\documentclass[10pt,final,twocolumn]{IEEEtran}

\usepackage{color}
\usepackage{subfigure}
\usepackage{graphicx}
\usepackage{amsmath}
\usepackage{amssymb}
\usepackage{amsfonts}
\usepackage{url}
\usepackage{algorithm}
\usepackage{algorithmic}
\usepackage{array}
\usepackage[T1]{fontenc}
\usepackage{paralist}
\usepackage{ulem}

\newcolumntype{M}[1]{>{\raggedright}m{#1}}

\def\imgratio{0.45}
\def\imgratiobis{0.55}

\newtheorem{theorem}{Theorem}
\newtheorem{proposition}{Proposition}
\newtheorem{lemma}{Lemma}

\renewcommand{\IEEEQED}{\IEEEQEDopen}

\begin{document}

\title{On-the-fly erasure coding for real-time video applications}

\author{
\IEEEauthorblockN{Pierre Ugo Tournoux$^{1,2}$, Emmanuel Lochin$^{1,2}$, J\'{e}r\^{o}me Lacan$^{2}$, Amine~Bouabdallah$^{1,2}$  and Vincent~Roca$^{3}$ \\}
\IEEEauthorblockA{$^1$ CNRS ; LAAS ; 7 avenue du colonel Roche, F-31077 Toulouse, France\\
$^2$ Universit\'{e} de Toulouse ; UPS, INSA, INP, ISAE ; LAAS ; F-31077 Toulouse, France\\
$^3$ INRIA, Plan\`{e}te research team, Grenoble, France}
}


\maketitle




\begin{abstract}




This paper introduces a robust point-to-point transmission scheme: Tetrys, that relies on a novel on-the-fly erasure coding concept which reduces the delay for recovering lost data at the receiver side.
In current erasure coding schemes, the packets that are not rebuilt at the receiver side are either lost or delayed by at least one RTT before transmission to the application. 
The present contribution aims at demonstrating that Tetrys coding scheme can fill the gap between real-time applications requirements and full reliability.
Indeed, we show that in several cases, Tetrys can recover lost packets below one RTT over lossy and best-effort networks. 
We also show that Tetrys allows to enable full reliability without delay compromise and as a result: significantly improves the performance of time constrained applications. 
For instance, our evaluations present that video-conferencing applications obtain a PSNR gain up to 7dB compared to classic block-based erasure codes.

\end{abstract}

\section{Introduction}
\label{sec:intro}
Multimedia applications, even over best effort networks, are more and more pervasive today.
This is the sign of an important need by end-users for such applications, no matter their location and the connection technology being used.
If the networking conditions are sometimes appropriate, users might also experience long transmission delays and significant packet losses.
When this happens, providing the level of data delivery timeliness and reliability required by multimedia applications seems to be really challenging \cite{milcom07}.
In this context, this work aims at providing a transport-level reliability mechanism, called Tetrys, compliant with real-time applications requirements and
able to recover lost packets in a given time threshold. 

Currently there are two kinds of reliability mechanisms based respectively on retransmission and redundancy schemes.
Automatic Repeat reQuest (ARQ) schemes recover all lost packets thanks to retransmissions.
This implies that the recovery delay of a lost packet needs at least to wait one supplementary Round Trip Time (RTT). However, this can be problematic 
if this delay exceeds the threshold of the application (i.e. the threshold above the application considers a packet outdated).

A well-known solution to prevent this additional delay is to add redundancy packets to the data flow. This can be done with the use of Application Level Forward Error Correction
(AL-FEC) codes\footnote{AL-FEC codes are FEC codes for the erasure channel where symbols (i.e. packets) are either received without any error or lost (i.e. erased) during transmission.}.
The addition of $n-k$ repair packets to a block of $k$ source packets allows to rebuild all of the $k$ source packets if a maximum of $n-k$ packets are lost among the $n$ packets sent.
In practice, only Maximum-Distance Separable codes (MDS), such as Reed-Solomon codes \cite{RFC5510}, have this optimal property, whereas other families of codes (like LDPC \cite{RFC5170} or Raptor codes \cite{raptor})
need to receive a few number of extra symbols in addition to the $k$ strict minimum.
However, if more than $(n-k)$ losses occur within a block, decoding becomes impossible.
In order to increase robustness (e.g. to tolerate longer bursts of losses), the sender can choose to increase the block size (i.e. the $n$ parameter) with the price of an increase of 
the decoding delay in case of erasure.
In order to improve robustness while keeping a fixed delay, the sender can also choose to add more redundancy while keeping the same block size with the price of a decrease of the goodput (which is not necessarily affordable by the application).
These trade-off between \textit{(packet~decoding~delay, block~length)} and \textit{throughput} are, for instance, addressed in \cite{flexibleRS09}.  
Another approach is proposed in \cite{martinian2004} where the authors use non-binary convolutional-based codes. They show that the decoding delay can be reduced with the use of a sliding window, 
instead of a block of source data packets, to generate the repair packets. However, both mechanisms do not integrate the receivers' feedbacks and thus, cannot provide any full reliability service.

Finally, an hybrid solution named Hybrid-ARQ which combines ARQ and AL-FEC schemes is often used.
This is an interesting solution to improve these various trade-off \cite{sahai:feedback08}.
However, when retransmission is needed, the application-to-application delay still depends on the RTT which might be not acceptable with real-time applications.

The present contribution totally departs from the above schemes. In fact, it inherits from the following two independent works on erasure coding which have converged
to an on-the-fly coding mechanism where feedbacks from the receivers are considered during the encoding process:

\begin{enumerate}
\item In \cite{medardIsit2008}, Sundararajan \textit{et al.} have proposed a coding scheme which includes feedback messages on the reverse path. The goal of this feedback path is 
to decrease the encoding complexity at the sender side without impacting on the communication transfer. This scheme allows to reduce the number of transmissions and as a result, 
the average decoding delay in the context of multiple receivers. In their evaluation, the authors neglect transmission delays and the resulting delays from the losses observed by 
different receivers. A noticeable contribution of their work is the concept of \textit{seen packet} by which the receiver acknowledges the degrees of freedom of the 
linear system corresponding to the received packets. This scheme has the main benefit of optimizing buffers occupancy while reducing the encoding complexity;
\item Independently, Lacan and Lochin also proposed in~\cite{lacan08iwssc} an on-the-fly coding system using feedbacks in the context of point-to-point communications with high transmission delays. 
Basically, the principle is to add repair packets generated as a linear combination of all the source data packets sent but not yet acknowledged. This scheme was proposed in order to enable 
full-reliability in Delay Tolerant Networks (DTN) and more specifically in Deep Space Networks (DSN) where an acknowledgment path might not exist and where the experienced delay might prevent the
efficient use of standard ARQ schemes. 
\end{enumerate}

Unlike current reliability methods, these on-the-fly coding schemes allow to fill the gap between systems without retransmission and fully reliable systems by means of retransmissions. 
In our work, we propose to deeply investigate the recovery delay of the lost packets, which is one essential characteristic of these on-the-fly coding schemes, and we show that
this delay is both tunable and independent of the RTT. 
The main contributions of this paper are the application of Tetrys \cite{tournoux09acmmul} (augmented with the concept of \textit{seen packets} \cite{medardIsit2008}) to the context of real-time
applications and the analysis of the performances achieved with a probabilistic approach.

We present the Tetrys mechanism in Section \ref{sec:proposal} and illustrate the simplicity of its configuration compared to FEC codes in Section \ref{sec:easyConfig}. 
Then we demonstrate in Section \ref{sec:video} that Tetrys offers significant gains compared to standard erasure coding schemes, in particular in terms of delay versus reliability trade-off
in the context of video-conferencing.
An exhaustive analytical study of the mechanism is given in Section~\ref{sec:models}.
It is followed by a performance analysis in Section~\ref{sec:redundancyAlloc}, that complements the experiments of Section \ref{sec:video}, and demonstrates that Tetrys is able to determine the
minimal amount of redundancy required to fulfill the application requirements.
We finally conclude this work in Section~\ref{sec:conclusion}.

\section{Proposal description}
\label{sec:proposal}
This section describes the Tetrys mechanism and the integration of the seen packet concept \cite{medardIsit2008}. We choose to introduce the main Tetrys principle in Section \ref{subsec:cannotwait} to allow the reader
a quick understanding of the present coding scheme used while Section \ref{subsec:details} further details Tetrys internal mechanisms. 

\subsection{Tetrys in a nutshell}
\label{subsec:cannotwait}

\begin{table}[htb!]
\begin{center}
\begin{small}
\begin{tabular}{|l|M{0.8\columnwidth}|}
\hline
$P_i$ & The $i^{th}$ source packet sent \tabularnewline
\hline
$R_{(i..j)}$ &  A repair packet built as a linear combination of the source packets $i$ to $j$: $R_{(i..j)}=\sum_{k=i}^{j}  \alpha^{(i,j)}_kP_k$ \tabularnewline
\hline
$k$ & The number of source packets between the transmission of two repair packets \tabularnewline
\hline
$n$ &  The total number of source plus repair packets for each group of $k$ source packets is denoted $n$ (to keep the usual definition) and is always equal to $k+1$ for Tetrys \tabularnewline
\hline
$R$ &  The redundancy ratio: $R=(n-k)/n=1-(k/n)$ where $k/n$ is the code rate. With Tetrys, we always have $R=1/(k+1)$ \tabularnewline
\hline
${\Delta}_{R}$ & The difference between the redundancy ratio and the packet loss rate: ${\Delta}_{R}=R-p$ \tabularnewline
\hline
$p$ & The packet loss rate (PLR) experienced \tabularnewline 
\hline
$b$ & The average burst size in case of a Gilbert Elliot channel. This value equals to $1$ in the particular case of a Bernoulli channel. Therefore this parameter also defines the type of erasure channel used \tabularnewline
\hline
$L_i$ & The $i^{th}$ lost packet \tabularnewline 
\hline
$F_{sack}$ & The feedback (i.e. acknowledgment) transmission frequency, at the receiver \tabularnewline 
\hline
$BS$ & The sender's (elastic) encoding window, composed of source packets not yet acknowledged \tabularnewline 
\hline
$BR$ & The receiver's buffer where the packets received and decoded are kept until they are no longer needed to decode \tabularnewline 
\hline
\end{tabular} 
\end{small}
\end{center}
\caption{Notations.}
\label{tab:notation}
\end{table}

Let us start with a quick overview of Tetrys.
The Tetrys sender uses an elastic encoding window buffer (denoted $BS$) which includes all the source packets sent and not yet acknowledged.
Let $P_i$ be the source packet with sequence number $i$.
Every $k$ source packets, the sender sends a (single) repair packet $R_{(i..j)}$, which is built as a linear combination (with random coefficients) of all the packets currently in $BS$.
The receiver is expected to periodically acknowledge the received or decoded packets. Each time the sender receives an acknowledgment, it removes the acknowledged packets from $BS$.
A receiver can decode lost packets as soon as the rank of the linear system, which corresponds to the available repair packets, is higher or equal to the number of lost packets.
In most cases, the decoding is successful as soon as the number of lost packets is lower or equal to the number of repair packets received. 

It results that: (1) Tetrys is tolerant to any burst of source, repair or acknowledgement losses, as long as the amount of redundancy exceeds the packet loss rate (PLR),
and (2) the lost packets are recovered within a delay that does not depend on the $RTT$, which is a key property for real-time applications.
These properties will be thoroughly studied in the remaining of this paper.

\subsubsection{A simple data exchange}

\begin{figure}[ht]
\begin{center}
\includegraphics[keepaspectratio=true, width=0.7\columnwidth]{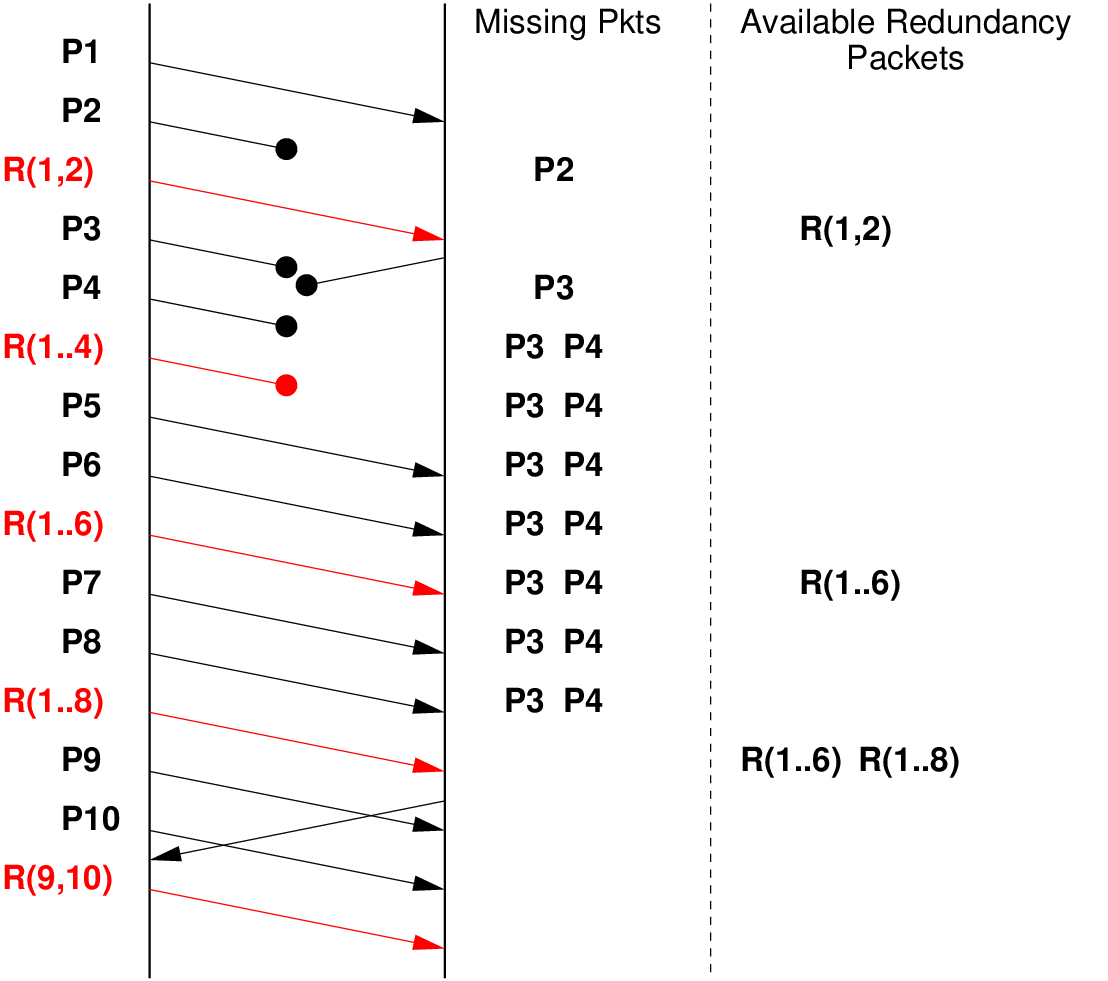}
\caption{A simple data exchange with Tetrys (k=2).}
\label{fig:ex1}
\end{center}
\end{figure}

Fig.~\ref{fig:ex1} illustrates a simple Tetrys exchange.
Here $k=2$ which means that a repair packet is sent each time two source packets have been sent.
The right side of this figure shows the list of packets that are lost and not yet rebuilt, as well as the repair packets kept by the receiver in order to recover them.
During this data exchange, packet $P_2$ is lost.
However, the repair packet $R_{(1,2)}$ successfully arrives and allows to rebuild $P_2$.
The receiver sends an acknowledgement for packets $P_1$ and $P_2$, in order to inform the sender that it can compute the next repair packets from packet $P_3$.
Unfortunately this acknowledgement is lost.
However this loss does not compromise the following transmissions and the sender simply continues to compute repair packets from $P_1$.
After this, we see that $P_3, P_4$ and $R_{(1..4)}$ packets are also lost.
These packets are rebuilt thanks to $R_{(1..6)}$ and  $R_{(1..8)}$ since the number of repair packets becomes higher or equal to the number of losses.

\subsection{A broader view of Tetrys}
\label{subsec:details}

We now detail the key concepts of Tetrys, namely the encoding and decoding process, the notion of seen packet, and the use of acknowledgments.

\subsubsection{Encoding process}
A repair packet is sent every $k$ source packets.
This packet is computed as a linear combination of all the source packets currently in $BS$, as follows:
	$$R_{(i..j)} = \sum_{l=i}^{j}  \alpha^{(i,j)}_l.P_l$$
\noindent where all packets between $P_i$ and $P_j$ belong to $BS$, with $\alpha^{(i,j)}_l$ are coefficients randomly chosen in a finite field $\mathbb{F}_{q}$, and where the multiplication of a coefficient by a packet is defined in \cite{Rizzo97a}.
From a practical point of view, instead of transmitting all the coefficients along with the associated repair packet (which introduces a potentially large transmission overhead), we use a Pseudo-Random Number Generator (or PRNG, e.g. \cite{randomGen}) and only transmit the seed which has been used.

The $k$ value is directly related to the code rate which is equal to $\frac{k}{k+1}$.
This is of course a key parameter that should ideally be adjusted dynamically depending on the network conditions.
For the sake of simplicity, the code rate is chosen fixed.
In section~\ref{sec:modelisation}, we analytically detail the code rate and evaluate with simulations its impact on the overall performance. We finally provide some guidelines to correctly set this value in Section~\ref{sec:redundancyAlloc}.

\subsubsection{Decoding process}

Decoding (i.e. recovering lost source packets) consists in solving the system of linear equations currently available at the receiver side.
The available source packets (received or decoded) are stored by the receiver as long as they might be used by the source to build the next repair packets $R_{(i..j)}$
while the repair packets are also stored as long as they can be used to recover lost packets.
More precisely, when a new repair packet $R_{(i..j)}$ arrives, all the available source packets that are part of {$P_i$ .. $P_j$} are subtracted from $R_{(i..j)}$.
The result is $R_{(L_1..L_l)}$, where $(L_1..L_l) \in (P_i..P_j)$ is the subset of packets of the linear combination that have been lost.

Let us assume that the $l$ source packets $(L_1..L_l)$ have been lost and that $l$ repair packets have been received and stored in $BR$.
Let $R^i$ be the $i^{th}$ packet of the set of $l$ repair packets (for the sake of readability, this notation does not mention the set of source packets used by the linear combination).
We obtain:
$$ (R^1,..,R^l)^T=G\cdot(L_1,.., L_l)^T $$
with:
\begin{equation}
G  = \left( \begin{array}{ccc}
\alpha^{R^1}_{L_1} & .. &\alpha^{R^1}_{L_l} \\
. & .. & .\\
. & .. & .\\
\alpha^{R^l}_{L_1} & .. &\alpha^{R^l}_{L_l} \\\end{array} \right)
\label{matrix1}
\end{equation}
and where $\alpha^{R^i}_{L_j}$ is the coefficient used to encode the $j^{th}$ lost packet in $R^i$.
If $G$ can be inverted, the lost packets $(L_1..L_l)$ are recovered with:
	$$ (L_1,.., L_l)^T=G^{-1}\cdot(R^1,..,R^l)^T $$
Once the decoding is successful, all of these $l$ repair packets can now be removed from $BR$.
If the matrix $G$ is singular, the repair packet whose coefficients are linearly dependent is discarded, and the receiver has to wait one more repair packet to do another attempt.

A solution to improve the probability of having an invertible matrix could consist in using super-regular matrices \cite{Hutchinson:superregular}.
However the dynamic nature of Tetrys makes this solution complex to set up.
Furthermore, it can be observed that with random coefficients, $G$ has an extremely high probability of being invertible if the finite field is chosen sufficiently large \cite{kahn2001}.

\subsubsection{Seen packet}
A lost packet is considered as "seen" by a receiver when it receives a fresh repair packet built from a linear combination that includes this lost packet (i.e. the lost packet was part of $BS$ at the time the repair packet has been created).
Even if a seen packet cannot be decoded immediately, the received repair packet contains enough information to recover this packet later. This explains why a "seen" packet acknowledges a 
source data packet as if it has been effectively received.
Of course, when several lost packets are covered by one repair packet, only the oldest lost packet is considered as seen.

\subsubsection{Acknowledgment packet}
A receiver periodically sends acknowledgment packets.
Each acknowledgment contains the list (in the form of a SACK vector \cite{rfc2018}) of the packets seen or effectively received or decoded.
Upon receiving this acknowledgment, the sender removes the acknowledged packets from the encoding window (BS).
Therefore these packets are no longer included in the linear combinations used to encode the next repair packets \cite{medardIsit2008}. This reduces the encoding/decoding complexity.
We choose to set the acknowledgment transmission frequency $F_{SACK}$, as a function of the current $RTT$: $F_{SACK} = s \times RTT$ where typical values for $s$ are ranging from $0.25$ to $2$ \cite{reducingtcpack}.
While the choice of $F_{SACK}$ does not impact on the reliability of the mechanism, there is a trade-off to find between the 
increase of $F_{SACK}$ which reduces the encoding/decoding complexity (evaluated in Section~\ref{sec:bufferSizing}) and the transmission overhead and acknowledgement processing cost.

\subsubsection{A complete example}

\begin{figure}[htb]
\begin{center}
\includegraphics[keepaspectratio=true, width=1.0\columnwidth]{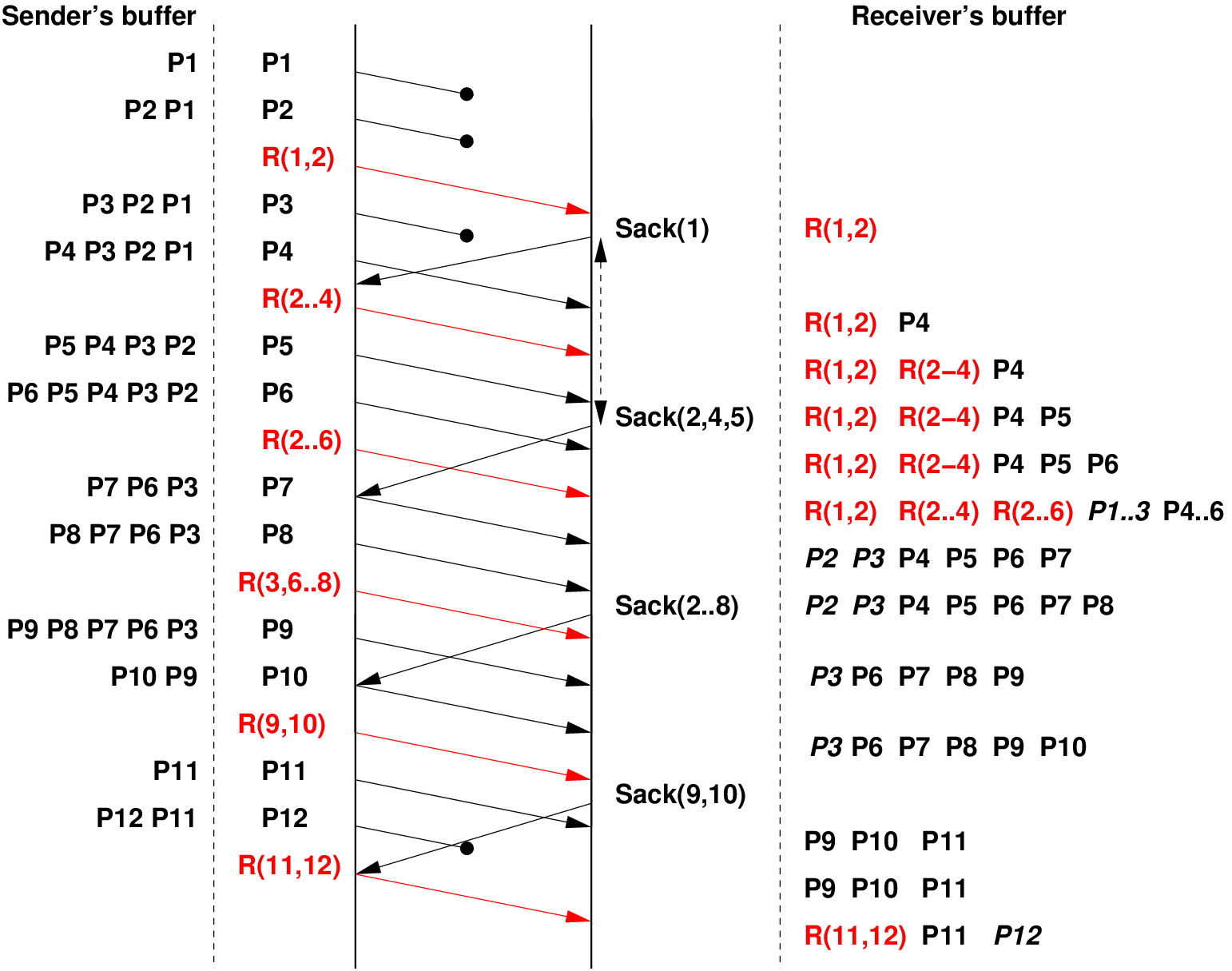}
\caption{A more elaborate data exchange, with selective acknowledgements and seen packets (k=2). Rebuilt packets are in italic.}
\label{fig:ex2}
\end{center}
\end{figure}

Let us consider the example of Fig.~\ref{fig:ex2}, where we assume the receiver sends back acknowledgments to a fixed frequency $F_{sack}$.
The sender first transmits packets $P_1$, $P_2$ and $R_{(1,2)}$.
Since the repair packet $R_{(1,2)}$ is the only one to be received, the receiver considers that $P_1$ and $P_2$ have been either lost or delayed.
Then, the receiver acknowledges packet $P_1$ since $R_{(1,2)}$ contains a linear combination of $P_1$ which is considered as "seen".
More generally, each time a repair packet is received, the receiver can acknowledge one of the source packets that are included in the linear combination.
Then, the sender transmits $P_3$ and $P_4$.
Just after, the sender receives an acknowledgement for packet $P_1$.
So the sender creates a new repair packet starting from $P_2$: $R_{(2..4)}$.
The receiver gets $P_4$ and $R_{(2..4)}$, meaning that the sender has received the previous SACK packet.
Then, the receiver sends a new SACK packet which acknowledges $P_2$, $P_4$, $P_5$.
The receiver cannot rebuild packets $P_1$ to $P_3$ since he did not receive enough repair packets.
As a result, the receiver stores $R_{(1,2)}$ and $R_{(2..4)}$ for a future use.
Since no loss occurs after that point, upon receiving a third repair packet, the receiver can now rebuild the missing packets.
The received source packets included in the linear combination are subtracted, which results in $R_{(1,2)},R^{\prime}_{(2..4)},R^{\prime}_{(2..6)}$ such as:

$$ (R_{(1,2)},R^{\prime}_{(2..4)},R^{\prime}_{(2..6)})^T=G\cdot(P_1,P_2,P_3)^T $$

with:

\begin{equation}
G  = \left( \begin{array}{ccc}
\alpha^{R_{(1,2)}}_{P_1} & \alpha^{R_{(1,2)}}_{P_2} & 0 \\
0 & \alpha^{R_{(2..4)}}_{P_2} & \alpha^{R_{(2..4)}}_{P_3}\\
0 & \alpha^{R_{(2..6)}}_{P_2} & \alpha^{R_{(2..6)}}_{P_3}\\\end{array} \right)
\label{matrix2}
\end{equation}

where $\alpha^{R_{(i..j)}}_{P_z}$ is the coefficient used to encode ${P_z}$ within the repair packet $R_{(i..j)}$.

With the assumption that $G$ is invertible, $G^{-1}$ is obtained thanks to a Gauss-Jordan elimination and packets $P_1$ to $P_3$ are given by:
	$$(P_1,P_2,P_3)^T=G^{-1}\cdot (R^{\prime}_{(1,2)},R^{\prime}_{(2..4)},R^{\prime}_{(2..6)})^T $$
These packets can be then considered as decoded. However, before removing them from $BR$, the receiver must still wait the reception of $R_{(3,6..8)}$ to be sure that the sender will not use these packets anymore to build new repair packets.

This example highlights the importance of several metrics: the decoding delay, the buffer size at the sender and at the receiver, and the number of operations needed to encode and decode. 
All these metrics will be studied and analyzed thoroughly in the Section \ref{sec:modelisation}.

\section{Evaluation of the Tetrys Parameters}
\label{sec:models}
This section includes both analytical and experimental evaluations of Tetrys.
To that purpose, we have implemented a Tetrys prototype in C language.
It borrows the finite field operations from Luigi Rizzo's Reed-Solomon codec \cite{Rizzo97a}.
For decoding, a Gauss-Jordan matrix inversion has been developed.
This algorithm is modified in order to determine, in the case of a singular matrix, the repair packet which is a linear combination of the other received packets.
This useless repair packet is then discarded and the decoder waits for additional repair packets.
During experiments, the coefficients for the linear combination are randomly chosen on the finite field  $\mathbb{F}_{256}$, except in Section \ref{sec:variableFiniteField} where other finite fields are used.

\subsection{Tetrys general analytical model}
\label{sec:modelisation}

We propose in this part a model allowing to assess the key properties of the Tetrys mechanism.
We assume the packet losses follow a Bernoulli law of parameter $p$. Under this assumption, we introduce a Markov chain: $\{Y_n,n>0\}$, which represents the difference 
between the number of lost packets and the number of received repair packets observed after the reception of each repair packet. As in section Sec.~\ref{sec:proposal}, we assume to decode when $Y_j=0$. 
This assumption is valid if the finite field is chosen sufficiently large (see \cite{kahn2001} for theoretical arguments and Section \ref{sec:variableFiniteField} for simulation results). 

As a first step, we focus on the probability distribution of $\{Y_n,n>0\}$. Then, we use this distribution to estimate the decoding delay, the average buffer size and the computation complexity of the algorithm.

The evaluation of $\{Y_n,n>0\}$ is done after each Tetrys block. We define a block as a set of $k+1$ consecutive packets that begins at the first source packet sent after a repair packet and ends at the next repair packet. We point out that our definition of block does not correspond to the usual definition in coding theory which is a set of symbols encoded together. In our context, a repair packet can be encoded from a set of source data packets belonging to several blocks. 

The reception of each packet is represented by a random variable (r. v.) $X_{i,j}$, where $i>0$ and $0\leqslant j \leqslant k$. With this notation, $i$ corresponds to the block and $j$ to the position of the packet in the block.

On the Bernoulli channel, we have $P[X_{i,j}=1]=p$ (the packet is lost), and  $P[X_{i,j}=0]=1-p$ (the packet is received). The variables $X_{i,j}$, where $0\leqslant j \leqslant k-1$ thus corresponds to source packets and the variables $X_{i,k}$ corresponds to the repair packets.   
We then define the r.v. $X_{i}$, where $i>0$, as follows:
\begin{equation}
\label{eq:bernoulli}
X_{i}=\sum_{j=0}^{k}X_{i,j}-1  
\end{equation}
Indeed, this sum can be expressed as $X_{i}=\sum_{j=0}^{k-1}X_{i,j} + (X_{i,k}-1)$. Then, the loss of one of the first $k$ (source) packet increments the value of $X_i$ while the reception of the repair packet decrements the value of $X_i$. Since $X_i$ is obtained from a sum of Bernoulli variables, we have $P(X_i=u-1)=\binom {k+1} {u}p^u(1-p)^{k+1-u} \text{~with~} u=0,\ldots,k+1$.   

We then define the Markov chain $\{Y_n,n\geqslant 0\}$ as follows: 
\begin{equation}
Y_n = \left\{ \begin{array}{ll}
Y_{n-1} + X_n & \textrm{if $Y_{n-1} + X_n \geqslant$ 0}\\
0 & \textrm{else}
\end{array} \right.
\end{equation}
Actually, the value of $Y_n$ corresponds to the difference between the number of lost packets and the number of received repair packets since the previous decoding. Note that this value is considered at the end of each block, {\textit i. e.} after the transmission of a repair packet.

\begin{theorem}
The success of the decoding and the decoding delay depend on the relationship between $R$ and $p$ as follows:
\begin{itemize}
\item if $R<p$, the recovery of a lost packet is not guaranteed;
\item if $R=p$, all the lost packets are recovered, but the mean decoding delay is infinite;
\item if $R>p$, all the lost packets are recovered, and the the mean decoding delay is finite;
\end{itemize}
\end{theorem}

\begin{IEEEproof}
From the definition of $X_i$, it can be shown that its expectation $E(X_i)$ is equal to $(k+1)p-1=\frac{p}{R} - 1$. If $R<p$, $E(X_i)$ is strictly positive and thus the chain is transient. Consequently, there is no guarantee to decode a lost packet.  

For $R=p$, $E(X_i)=0$ and the chain becomes null recurrent, i. e. any state can be reached, but in an infinite time. Since the state $0$ corresponds to a decoding, it can be deduced that any lost packet is decoded but the mean decoding delay is infinite.

For $R>p$, $E(X_i)<0$ and thus the state $0$ is positive recurrent. This state is reached in a finite mean time and thus any lost packet is decoded in in finite decoding delay.
\end{IEEEproof}

Let us consider the case where $R>p$. Before studying the decoding delay in the next part, we can deduce additional informations on the decoding process from the Markov chain. Let us denote  $a_{i,j}:=P(Y_n=j|Y_{n-1}=i)$ the transition probabilities between the states $i$ and $j$. Let us now define $A$ the matrix $(a_{i,j})_{i,j\geqslant 0}$ and let us denote by $a_{i,j}^{(n)}$ the entries of $A^n$.

\begin{proposition}
If $R>p$, the chain $\{Y_n,n\geqslant 0\}$ admits a stationary distribution equal to :
\begin{equation}
P(Y_j=i)= \lim_{n \to \infty} a_{j,i}^{(n)}
\end{equation}
for any $i,j\geqslant 0$.
\end{proposition}

\begin{IEEEproof}
Since the chain is irreducible and one state is positive recurrent, all the states are positive recurrent \cite{CoxDavid1965Theory}. Thus the chain admits a stationary distribution whose values can be easily obtained with basic results in stochastic process theory \cite{CoxDavid1965Theory}.
\end{IEEEproof}


\subsection{Analytical model of the decoding delay}
\label{subsec:decodingdelay}

To study the decoding delay, we first need to obtain the distribution of the first hitting time. In our context, the first hitting time is denoted by $H_{i}$ and is defined as follows:
$$H_{i}=\{\min h \textrm{ such that } Y_h=0 | Y_0=i  \}$$
Intuitively, this hitting time corresponds to the time necessary to decode a packet knowing that, at the considered time, the difference between the number of lost packets and the number of received repair packets is $i$.  

\begin{lemma}
The probability distribution of $H_{i}$ can be obtained as follows :
\begin{equation}
P(H_{i}=h)=\frac{1}{h!}\frac{d^h (\sum_{t\geqslant 0} a_{i,0}^{(t)}z^t/\sum_{t\geqslant 0} a_{0,0}^{(t)}z^t)}{d z^h}|_{z=0}
\end{equation}
\end{lemma}

\begin{IEEEproof}
Let us define 
\begin{equation}
\label{eq:Gi}
G_{i}(z)=\sum_{t\geqslant 0} a_{i,0}^{(t)}z^t
\end{equation} 
and 
\begin{equation}
\label{eq:Fi}
F_{i}(z)=\sum_{h\geqslant 0} P(H_{i}=h)z^h
\end{equation}
\noindent the probability generating function (p. g. f.) of $H_{i}$. Following \cite[chap. 2, lemma 25]{aldous-fill:book}, we have :
\begin{equation}
\label{eq:Fi2}
F_{i}(z)=G_{i}(z)/G_{0}(z)
\end{equation}

The probability distribution of $H_{i}$ can be then obtained from the probability generating function by evaluating:
\begin{equation}
\label{eq:HiF}
P(H_{i}=h)=\frac{1}{h!}\frac{d^h F_{i}(z)}{d z^h}|_{z=0}.
\end{equation}
Combining Equations \ref{eq:Gi}, \ref{eq:Fi2} and \ref{eq:HiF} allows to obtain the expression of the probability distribution of $H_{i}$.
\end{IEEEproof}

Since this Markov chain concerns the decoding delay at the block level, we now need to refine the analysis at the packet level.  Let us consider that a packet sent in position $j$ ($j=0,\ldots,k-1$) of a block $i$ is lost. Let $D_j$ be its decoding delay. This delay has necessarily the form $k-j+h(k+1)$ because the decoding can only be performed at the reception of a repair packet. 

\begin{proposition}
The decoding delay of a packet sent in position $j$ of a block has the following distribution :
\begin{equation}
\begin{small}
\begin{array}{l}
P(D_j=k-j+h(k+1))  =\\
\sum_{y\geqslant 0} \sum_{u=0}^{k} \binom{k}{u}p^u(1-p)^{k-u}  P(H_{y+u}=h) P(Y_{i-1}=y)  \\
\end{array}
\end{small}
\end{equation}
\end{proposition}
\begin{IEEEproof}
Recall that $Y_{i-1}$ and $Y_{i}$ are the r. v. representing the states of the chain $\{Y_n,n\geqslant 0\}$ after the previous block and at the end of the current block. 

Since the packet sent in position $j$ is lost, we have:
\begin{equation}
\label{eq:YiYi-1}
P(Y_i=y+u|Y_{i-1}=y) = \binom{k}{u}p^u(1-p)^{k-u}
\end{equation}
\noindent for $u=0,\ldots,k$. We also have:
\begin{equation}
\begin{array}{l}
P(D_j=k-j+h(k+1))  \\
\  =\sum_{y\geqslant 0} \sum_{u=0}^{k} P(D_j=k-j+h(k+1),\\ 
\hspace{1.5cm} Y_{i-1}=y,Y_{i}=y+u) \\ 
 =\sum_{y\geqslant 0} \sum_{u=0}^{k} P(D_j=k-j+h(k+1)|Y_{i}=y+u) \\ 
\hspace{1.5cm} P(Y_{i}=y+u|Y_{i-1}=y) P(Y_{i-1}=y)  \\
=\sum_{y\geqslant 0} \sum_{u=0}^{k} P(H_{y+u}=h) \\
\hspace{1.5cm} P(Y_{i}=y+u|Y_{i-1}=y) P(Y_{i-1}=y)  \\
\end{array}
\end{equation}
Combining this last expression with Equation \ref{eq:YiYi-1} allows to obtain the expression given in the proposition.
\end{IEEEproof}

\subsection{Analytical model of the matrix sizes}
\label{sec:matrixSize}
Like most of erasure codes, the decoding operation in Tetrys basically consists in inverting a matrix  defined over a finite 
field. The size of this matrix, denoted by $Z$, corresponds to the number of repair packets involved in the decoding. Compared to classic block-based erasure codes (rateless or not), the main difference 
is that no theoretical bounds exist on
the size of the matrix that must be inverted. This is due to the concept of elastic coding window. On the other hand, thanks 
to the elastic coding window, it can be observed that, with a good choice of parameters, the sizes of the inverted matrices 
by Tetrys is most of the time lower than the matrices used by classic erasure codes. For these reasons, the study of the 
sizes' distribution of the inverted matrices is important.  

The first step in this study is the analysis of the recurrence time. This parameter, denoted by $U$, is the time between the first loss after a decoding  and its recovery. This time is expressed in time units, where a unit time corresponds to the delay between the transmission of two consecutive packets.

With the notations introduced in the previous section, if we consider the block where the first packet is lost after a decoding, we define the r. v. $F$ which corresponds to the position 
of the first lost packet in the block.  When the first lost packet occurs in position $j$, its recovery delay, and thus the corresponding recurrence time $U$ has the form  $k-j+h(k+1)$, where $h$ represents the number of complete blocks included in the recurrence time. 
Reciprocally, a recurrence time equal to $k-j+h(k+1)$ can only be observed with a first loss at position $j$.

\begin{lemma}
\label{eq:PU1}
The recurrence time $U$ has the following distribution :
\begin{equation}
\begin{small}
\begin{array}{l}
P(U=k-j+h(k+1))  =\\
\hspace{0.5cm} \frac{1}{1-(1-p)^k}\sum_{u=0}^{k} \binom{k-j}{u}p^{u+1}(1-p)^{k-u}P(H_{u}=h)\\
\end{array}
\end{small}
\end{equation}
\end{lemma}

\begin{IEEEproof}
Basic combinatorial arguments show that 
\begin{equation}
\label{eq:PF}
P(F=j)=p(1-p)^j/(1-(1-p)^k), 
\end{equation}
for $j=0,\ldots,k-1$.

Since the considered packet is the first lost after the previous decoding, the value of the next $Y_i$ is necessarily in the range $[0,k]$. 
Thus, we have:  
\begin{equation}
\begin{small}
\begin{array}{l}
P(U=k-j+h(k+1)) =\\
\sum_{u=0}^{k} P(U=k-j+h(k+1),Y_{i}=u|F=j)P(F=j)
\end{array}
\end{small}
\end{equation}
It follows that:
\begin{equation}
\label{eq:PU2}
\begin{array}{l}
P(U=k-j+h(k+1)) \\
 = \sum_{u=0}^{k} P(D_j=k-j+h(k+1)|Y_{i}=u) \\ 
\hspace{1.5cm} P(Y_{i}=u|F=j)P(F=j)  \\
= \sum_{u=0}^{k} P(H_{u}=h)P(Y_{i}=u|F=j)P(F=j) \\
\end{array}
\end{equation}
It can easily be shown that $P(Y_{i}=u|F=j)= \binom{k-j}{u}p^u(1-p)^{k-j-u}$. By combining this result with Equations \ref{eq:PF} and \ref{eq:PU2}, we obtain the probability distribution of $U$ given in the lemma.
\end{IEEEproof}
\begin{proposition}
The distribution probability of $Z$, representing the sizes of the decoded matrices, is equal to:
\begin{equation}
\begin{small}
\begin{array}{l}
P(Z=i) =  \frac{1}{1-(1-p)^k} \sum_{h\geqslant i}\sum_{j=0}^{k-1} \sum_{u=0}^{k} \binom{h}{i} \binom{k-j}{u} \\
\hspace{3.5cm} p^{h-i+u+1}(1-p)^{i+k-u}  P(H_{u}=h)
\end{array}
\end{small}
\end{equation}
\end{proposition}
\begin{IEEEproof}
To obtain the matrix size $Z$ from $U$, we can first observe that in a recurrence time equals to $k-j+h(k+1)$, $h+1$ repair symbols are sent. This means that the matrix size is ranging 
from $1$ to $h+1$. By considering that the last repair symbol is necessarily received, we have:
\begin{equation}
\label{eq:Z|U}
P(Z=i|U=k-j+h(k+1))=\binom{h}{i}p^{h-i}(1-p)^i 
\end{equation}

On the other hand, we have:
\begin{equation}
\begin{array}{l}
P(Z=i) = \sum_{h\geqslant i}\sum_{j=0}^{k-1} P(Z=i|U=k-j+h(k+1)) \\
\hspace{3cm} P(U=k-j+h(k+1))
\end{array}
\end{equation}
By combining this expression with Equations \ref{eq:PU1} and \ref{eq:Z|U}, we obtain the given formula. 
\end{IEEEproof}

\subsection{Analytical model of the buffer size}
\label{sec:bufferSizing}
Like for the matrix sizes, the elastic coding window of Tetrys implies that there is no theoretical bounds on the number of packets stored in the buffer at the sender and  receiver sides. The aim of this part is to evaluate these parameters. In this section, we consider that a packet is sent by the sender each time unit.  

\subsubsection{At the sender side}

We denote by $BS_t$ the number of packets stored in the buffer at time $t$. Basically, the buffer contains the packets that were not acknowledged. Let $S_1$ denotes the time between the reception of the last SACK and $t$. If we consider that a SACK is sent every $s.RTT$ time units and that it is lost with probability $p$, we have :
\begin{equation}
E(S_1)= s.RTT(1/2 + 1/(1-p))
\end{equation}
The factor $1/2$ corresponds to the average time to wait a received acknowledgment and the factor $1/(1-p)$ is the expectation of the geometrical law of parameter $p$ representing the arrival of the last SACK.

This acknowledgment brings out the information on the reception of the packet sent by the sender one $RTT$ ago. Thus, the sender has to store the $RTT.k/(k+1)$ source packets sent during this period. 

Finally, at the time $t-S_1-RTT$, some source packets were not acknowledged because they were lost. Thanks to the use of the \textit{ack-when-seen} mechanism (included in the SACK mechanism), each received repair packet acknowledges a lost source packet. Thus, the number of not acknowledged source packets is the difference between the number of lost source packets and the number of received repair packets, which is represented by the r. v. $Y_n$ studied in Section \ref{subsec:decodingdelay}. 

The average number of packets stored in the buffer is thus:
\begin{equation}
\label{eq:meanBSt}
E(BS_t) =  RTT(k/(k+1))(s/2 + s/(1-p)) + E(Y_n)
\end{equation}
%

Since the RTT does not impact on the value of $E(Y_n)$, we can observe that, when we fix the other parameters ($p$, $k$ and $s$), the number of packets in the sender buffer is a linear function of the RTT. This observation also holds for the parameter $s$ representing the SACK frequency.

\subsubsection{At the receiver side}

The receiver has two buffers: the source buffer, which contains the received source packets necessary for future decoding and the repair buffer, which contains the received repair packets not yet decoded. The number of packets in the source buffer at the time $t$ is denoted $BRS_t$ and the number of packets in the repair buffer is denoted $BRR_t$.

We recall that, when a source packet is received by the receiver, it is acknowledged in the future SACKs. When the sender received the first of these SACKs, it deletes this source packet in its buffer and does not include it in the generation of the next repair packets. The receiver can delete this source packet as soon as it received a repair packet which does not include this source packet in its linear combination.

\begin{figure}[htb]
\begin{center}
\includegraphics[width=0.5\columnwidth]{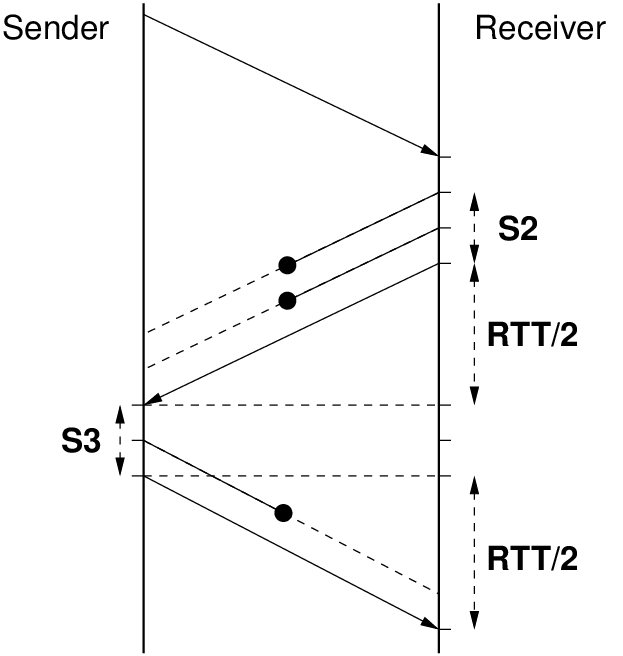}
\caption{Receiver buffer}
\label{fig:receiver_buffer}
\end{center}
\end{figure}

As shown in Fig.~\ref{fig:receiver_buffer}, it follows that the source packet is stored in the buffer during $S_2+S_3+RTT$, where $S_2+RTT/2$ is the time needed by the sender to receive the first acknowledgment and $S_3+RTT/2$ is the time needed by the sender to receive the next repair packet. 

Clearly, $S_2$ follows the same law than $S_1$. For $S_3$, the same method can be used to estimate the mean, excepted that a repair packet is sent each $k+1$ time units (instead of $s.RTT$ for the SACKs). 

The average time spent by a source packet in the buffer is then:

$$E(S_2+S_3+RTT)= RTT+(k+1+s.RTT)(1/2 + 1/(1-p))$$

To obtain the number of packets stored in the buffer at a given time, we must consider that some of these packets are lost. Thus we have:

\begin{displaymath}
\begin{array}{ll}
E(BRS_t) & = (k/(k+1))(1-p)E(RTT+S_2+S_3) \\
         & = (k/(k+1))(1-p)RTT+(k+1+s.RTT)\\
         & ((1-p)/2 + 1)
\end{array}
\end{displaymath}

To estimate the number of repair packets in the repair buffer, we can first estimate the probability of having no repair packet in the buffer. This probability is equal to $P(Y_n=0)$ determined in Section \ref{subsec:decodingdelay}.
   
When there is at least one packet in the repair buffer, we can consider the probability distribution of the recurrence time $U$. Indeed, for $U=k-j+h(k+1)$, $h$ repair packets are sent and we can estimate that, on average, $(1-p)h$ repair packets are received. It follows that the average number of packets in the buffer during this period is $(1-p)h/2$.
We then have:
\begin{displaymath}
\begin{array}{l}
E(BRR_t) = \frac{ \sum_{ h>0} (1-p)h \sum_{j=0}^{k-1} (k-j+h(k+1))P(U=k-j+h(k+1)) }{2.P(Y_n=0)}
\end{array}
\end{displaymath}

Following this model, we can assess the minimum buffer size requested by Tetrys. In addition, source-based algorithms can also be envisaged to prevent buffer overflow.

\subsection{Experimental evaluation of the buffer size}

\begin{figure}
\begin{center}
\subfigure[Number of data source packets in the receiver's buffer as a function of the acknowledgement frequency for a PLR=0.1]{
\includegraphics[keepaspectratio=true, width=6cm]{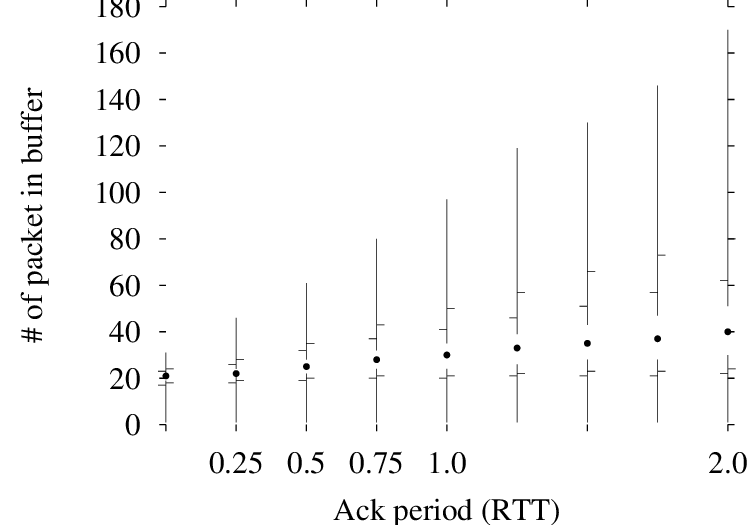}
\label{fig:ackFrequency}
}
\subfigure[Number of source data packets in the receiver's buffer as a function of the PLR for one acknowledgement sent per RTT]{
\includegraphics[keepaspectratio=true, width=6cm]{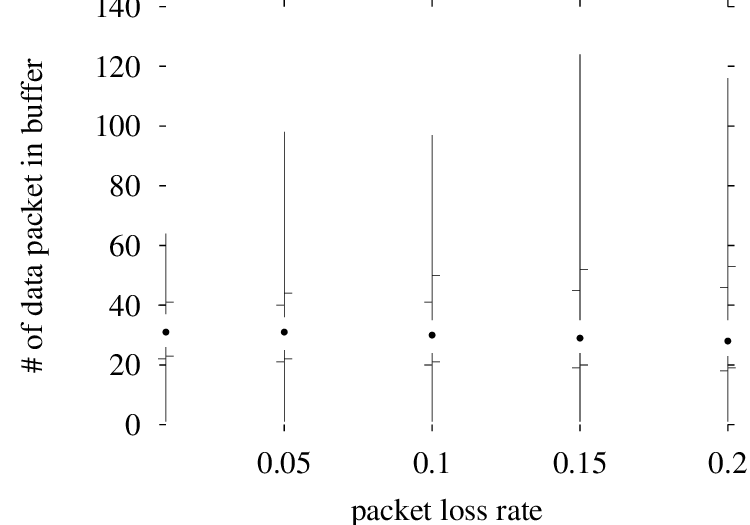}  
\label{fig:Buffer0Plr}
}

\caption{Minimum, maximum and (5, 10, 25, 50, 75, 90, 95) percentiles of the number of packets requested to decode with a 3/4 repair ratio for Tetrys}
\label{fig:simu-buffer}
\end{center}
\end{figure}

In order to give an insight of the Tetrys requirements in a typical case, we evaluate the data source receiver buffer ($BRS_t$) evolution using our Tetrys prototype.
We report only experiments over a Bernoulli channel\footnote{The results are in the same order of magnitude with bursty losses, using a Gilbert-Elliott channel.} for the receiver's buffer as the receiver's buffer occupancy is always bigger than the sender. The RTT, repair ratio and sending rate are respectively set to $200ms$, $(3/4)$ and $100$ packets per seconds. The two parameters that might affect the requested buffer sizes are the acknowledgment frequency (as presented Section \ref{sec:proposal}) and the PLR. We studied in Fig.~\ref{fig:ackFrequency} the impact of the acknowledgment frequency on the requested buffer size. Experiments are done with a fixed loss rate (10\%). For the sake of completeness, we show the minimum, maximum and the (5, 10, 25, 50, 75, 90, 95) percentiles (the 50 percentile is the buffer size of the 50\% highest buffer sizes) of the number of packets in buffer during the experiment. The samples used to compute these percentiles are selected at the reception of each data or repair packets.  

We can see that with one acknowledgment sent per packet, one per RTT and one for two RTT the $50$th percentile are respectively around $20$, $30$ and $40$ packets. The points in Fig.~\ref{fig:ackFrequency} also give the mean value which overlaps the $50$th percentile. This confirms that as $E(BRS_t)$ suggests, the average number of packets kept in the buffer evolves linearly with the acknowledgment frequency.

The other parameter of interest is the PLR, since we have seen that when its value is closed to the repair ratio, the recurrence time increases. Fig.~\ref{fig:Buffer0Plr} presents the result with 
an acknowledgment frequency of $1$ and shows the number of packets in the buffer for a PLR varying from 1\% to 20\%. We can see that the (5, 10, 25, 75, 90, 95) percentiles remain close to their 50 percentile, implying a low number of packets in the buffer (most of the time around 30 $\sim$ 40 for one acknowledgement per RTT) and a reasonable peak size (a maximum of 160 packets) the rest of the time.


\subsection{Tetrys encoding/decoding complexity analysis}
\label{sec:complexity}

This section introduces a complexity analysis of Tetrys operations, expressed in terms of the number of operations performed on packets.
For example, the multiplication of a packet by a finite field coefficient or the XOR addition of two packets are considered as one operation. 

\subsubsection{Encoding Complexity}
This complexity corresponds to the number of operations needed to generate one repair packet. Following the main principle of Tetrys, the number of source packets involved in the linear combination is the number of packets not acknowledged, i.e. the number of source packets in the buffer of the sender. The number of additions and multiplications performed to generate a repair packet at time $t$ is exactly $BS_t$. An analytical expression of this parameter is given in Equation \ref{eq:meanBSt}. Following discussions of Section 
\ref{sec:bufferSizing}, for fixed packet loss rate and redundancy ratio, this complexity is linear according to the RTT and to the SACK frequency.

\subsubsection{Decoding Complexity}
The decoding process can be split into two separate processes. The first one is a continuous process which consists in subtracting all the available source packets (received or decoded) to the repair packets in which they are involved. The second one is the core decoding process which allows to recover a set of $Z$ source packets from a set of $Z$ repair packets. As explained before, the $Z\times Z$-matrix built from the finite field symbols used to generate the repair symbols is inverted and the obtained matrix is multiplied to the vector of repair symbols to recover the source symbols.

To evaluate the complexity of the first process, it is sufficient to estimate the number of available source packets in the source buffer of the receiver. This quantity, $BRS_t$, is studied in Section \ref{sec:bufferSizing}. Figures \ref{fig:ackFrequency} and \ref{fig:Buffer0Plr} confirm these results with simulation results showing the evolution of the buffer size, and thus of this complexity, for typical parameters.

For the second process of the decoding operation, the decoder has to invert a matrix of size $Z$  and then to multiply the $Z\times Z$-inverted matrix by the vector of $Z$ repair packets. The matrix-vector multiplication only perform $Z$ operations on each repair packets. The inversion of a general matrix has a cubic complexity, but it is done on finite field coefficients and not on packets. In practical, when the entries of the matrix are carefully chosen, it can be shown that this matrix inversion does not strongly impact on the decoding speed for moderate values of $Z$.

The distribution of the parameter $Z$ was analytically studied in Section \ref{sec:matrixSize} for the Bernoulli channel. Simulation results obtained for typical parameters perfectly fit these theoretical estimations (see Figure \ref{fig:varp}). For a Gilbert-Elliott (GE) channel, additional simulations presented on Figure \ref{fig:varb} show the behavior of the $Z$ parameter on bursty channels. 

To have roughly estimations of the practical decoding speed, Tetrys decoding can be compared to a block code decoding with dimension equal to $Z$. 
In Fig. \ref{fig:varp} and \ref{fig:varb}, the highest average matrix size is equal to $14$. As a result, we can compare the cubic complexity of the matrix  inversion process to an erasure code of equivalent dimension defined over a non-binary finite field such as Reed-Solomon. If we now consider the subtraction process of source symbols from redundancy packets, Tetrys could be compared to common Reed-Solomon code of dimension $32$ (assuming the source data buffer size from Fig.~\ref{fig:Buffer0Plr}). To roughly have an order of magnitude, the authors in~\cite{ccncSoro} show that several implementations of Reed-Solomon code of dimension $32$ can reach a decoding speed up to $600$ Mbps with a standard personal computer. As a result, Tetrys is perfectly compliant with real-time video constraints both in terms of computation overhead and memory footprint 
(also practically observed with our real prototype).



\subsection{Experimental analysis of the impacts of the channel type}
\label{sec:simulations}

\subsubsection{Case of a Bernoulli channel: impact of the PLR}
\label{sec:inputProbas}

\begin{figure}[ht]
\begin{center}
\includegraphics[keepaspectratio=true, width=\imgratiobis\columnwidth, angle=-90]{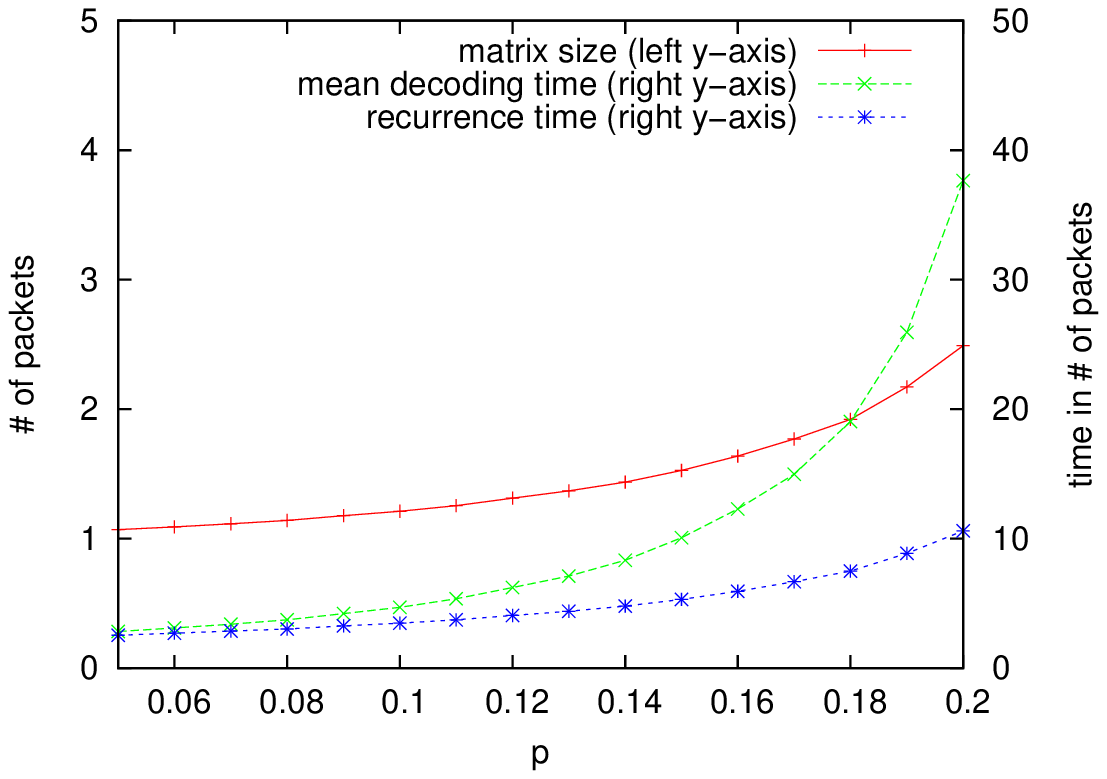}
\caption{Average matrix size, decoding delay and recurrence time as a function of the PLR, $p$, using a Bernoulli channel model.}
\label{fig:varp}
\end{center}
\end{figure}

We first consider the impact of the PLR, $p$, using a simple Bernoulli channel model, on Tetrys performance.
In Fig.~\ref{fig:varp}, the Tetrys performance in terms of average matrix size, decoding delay, and recurrence time is illustrated as a function of the PLR, using a Bernoulli channel, when $R=0.25$.
The first y-axis scale (left side) is expressed in number of packets and is used for the average matrix size.
The second y-axis scale (right side) is expressed in time units and is used for the average decoding delay and average recurrence time (recall that a time unit corresponds to the delay between the transmission of two consecutive packets).

The first observation is that the three curves increase with the PLR.
This is easily explained by the fact that when the error probability is small compared to $R$, then decoding happens quickly, and vice-versa.
This is also in line with a previous result showing that he average recurrence time is equal to $1/(R-p)$ and thus, is infinite when $R=p$.
The second observation is that the average decoding delay curve gets higher than the recurrence time curve.
This can be explained by the fact that the decoding delay is related to packet while the recurrence time is related to decoding.
In the case of a large ``recurrence walk'', a large number of packets have a large decoding delay, and thus this walk has a larger influence on the average decoding time than on the average recurrence time.

\subsubsection{Case of a Gilbert Elliot channel: impact of the average loss burst size}
\label{sec:variableBurstSize}

\begin{figure}[htb]
\begin{center}
\scalebox{0.4}{ \input{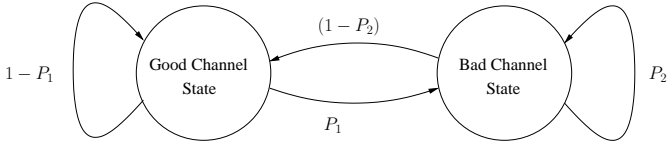} }
\caption{The first-order two-state Markov chain representing the Gilbert-Elliott channel model}
\label{fig:goodbad}
\end{center}
\end{figure}

\begin{figure}[ht]
\begin{center}
\includegraphics[keepaspectratio=true, width=\imgratiobis\columnwidth, angle=-90]{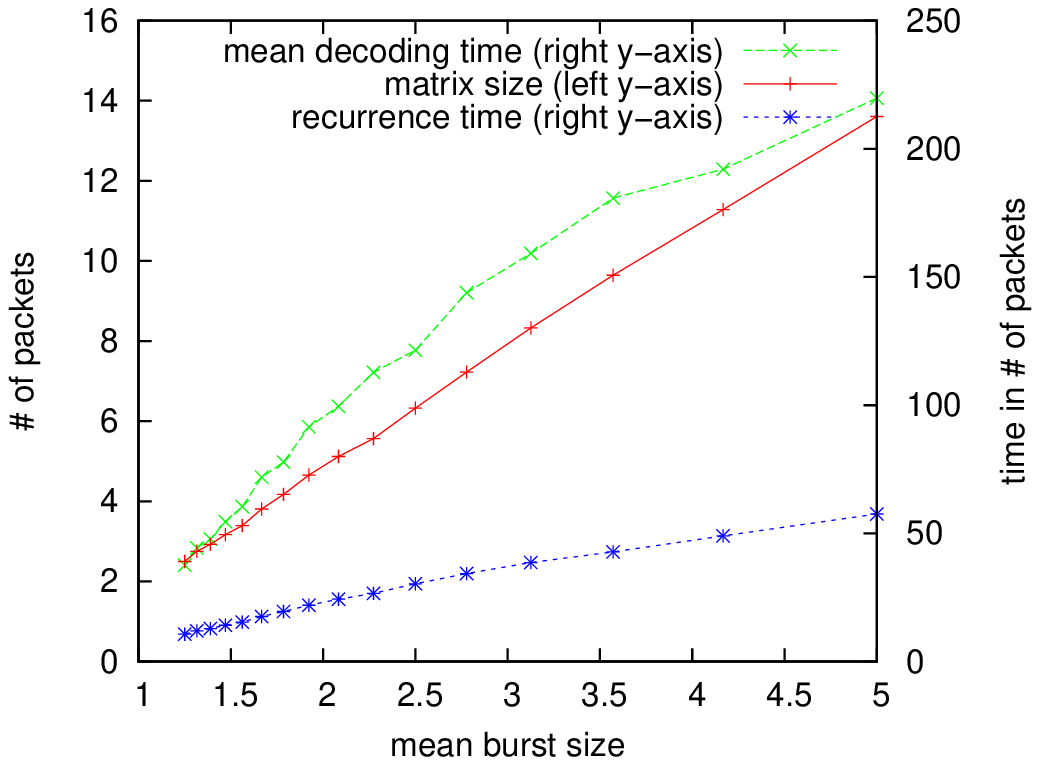}
\caption{Average matrix size, decoding delay and recurrence time as a function of the average loss burst size, using a GE channel model.}
\label{fig:varb}
\end{center}
\end{figure}

We now consider the impact of loss bursts on Tetrys performance, using the well-known first-order, Gilbert-Elliott channel model (Fig.~\ref{fig:goodbad}).
With this model, which considers two input probabilities, $p_1$ and $p_2$, it is well known that the mean PLR is equal to $p=p_1/(1+p_1-p_2)$ and the average loss burst size to $1/(1-p_2)$.
Thus: $p_2 = 1 + p_1 - p_1/p$. 

Fig.~\ref{fig:varb} shows the Tetrys performance, using the same metrics as before, as a function of the average loss burst size, when $R=0.25$.
During the tests, $p_1$ and $p_2$ vary in such a way that the mean PLR is kept constant, equals to $0.2$.

Compared to Fig.~\ref{fig:varp}, the curve representing the average loss burst size (equal to $1/(1-p_2)$) is added. We can observe that a small value of $p_1$ implies a large value of $p_2$ and thus a large mean burst size. On the opposite, when $p_1=p_2$, the Markov channel becomes a Bernoulli channel of parameter $p_1$ and thus, the mean burst size reaches its minimum.

The main information of the Fig.~\ref{fig:varb} is that the burst losses have a negative impact on Tetrys performance.
We can observe that when $p_1$ varies from $0.1$ to $0.2$, the burst size varies from $2.5$ to $1.25$. In this range, the matrix size, mean decoding time and recurrence time are also divided by 2.

Even this rate of 2 is very specific to this simple example, more generally, we can observe that the only consequence of bursts is the increase of the decoding delay, recurrence time and of the matrix size at the decoder side. 
Indeed, the property to decode all packets if $R>p$ remains true.

Note that in the case of channels with variable parameters (with a fixed PLR), Tetrys adapts automatically to the variable conditions without any external intervention.

\subsection{Experimental analysis of the impact of the finite field size}
\label{sec:variableFiniteField}

\begin{figure}[ht]
\begin{center}
\includegraphics[keepaspectratio=true, width=\imgratiobis\columnwidth, angle=-90]{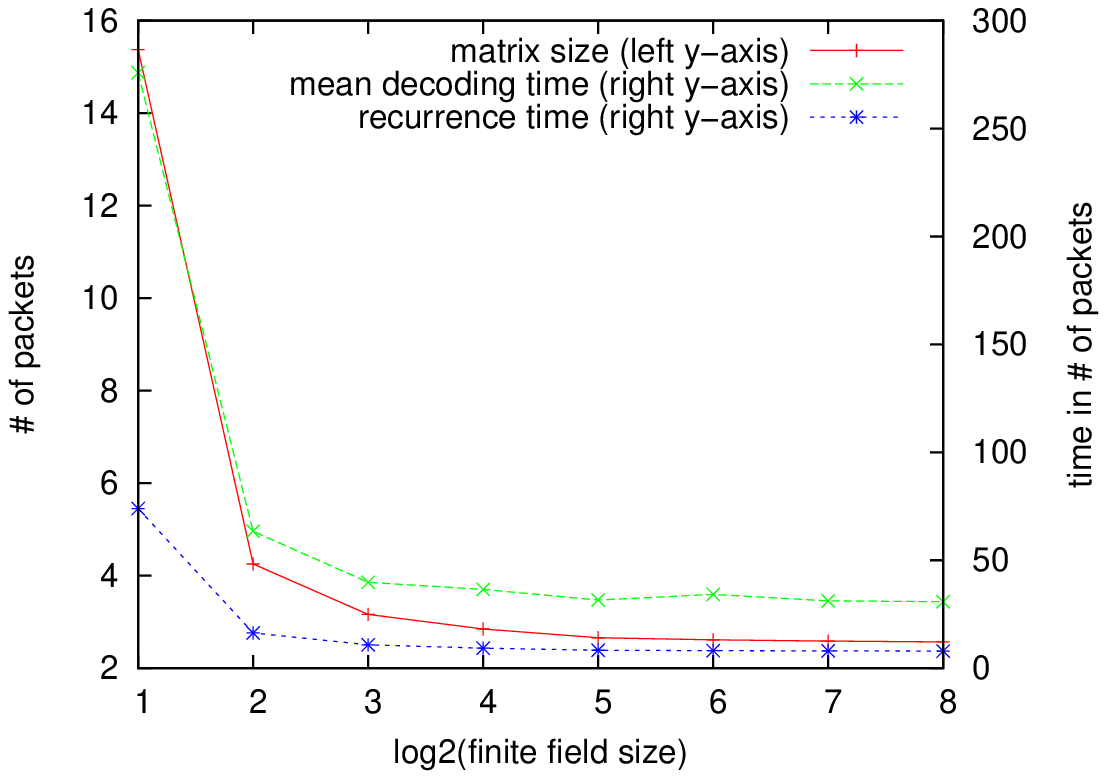}\hfill
\includegraphics[keepaspectratio=true, width=\imgratiobis\columnwidth, angle=-90]{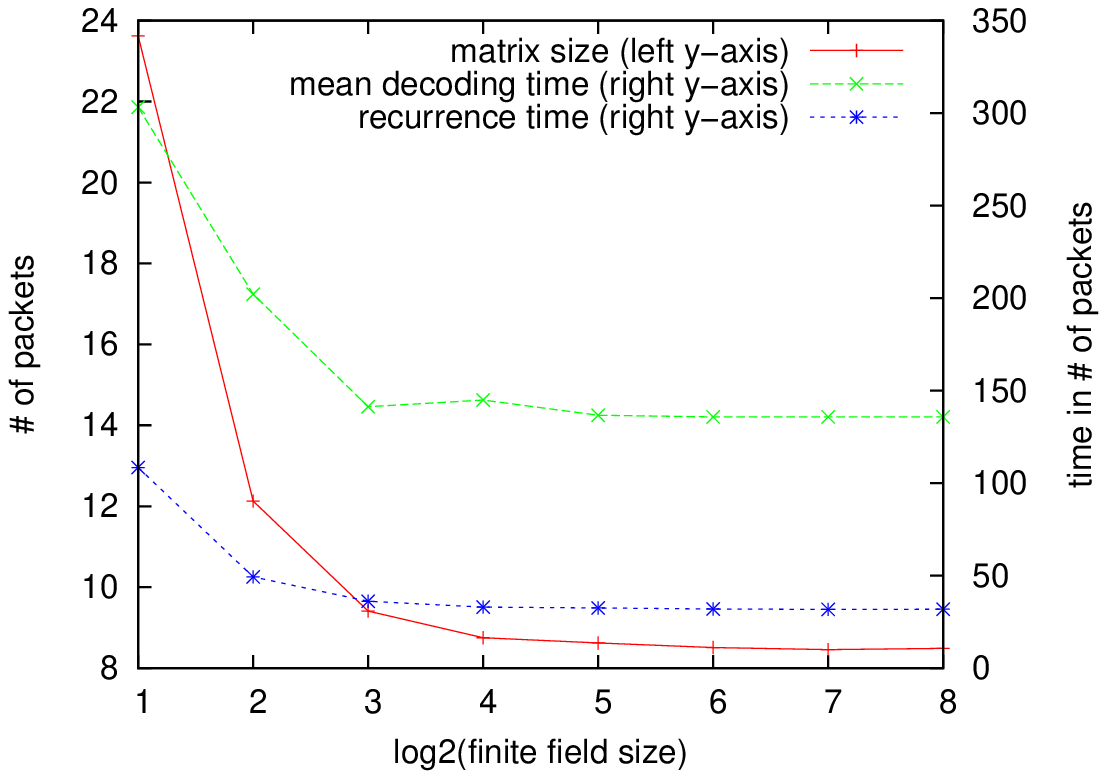}
\caption{Impact of the finite field size on the average matrix size, decoding delay and recurrence time, using a Bernoulli channel (top) or Gilbert Elliott channel (bottom). PLR=0.2, average loss burst size of 3 (GE channel case), and R=0.25.}
\label{fig:varf}
\end{center}
\end{figure}


Section \ref{sec:proposal} says that decoding is not necessarily possible as soon as the number of received repair packets is equal to the number of lost source packet. This is explained by the fact that the corresponding matrix can be singular (i.e. non invertible). In this case, the receiver must wait additional repair packets, which increases both the decoding delay and the matrix size.
In this section we analyze the impacts of the finite field size (over which the coefficients used to build the repair packets are randomly chosen) on these performance metrics.

More precisely we carried out experiments where the finite field size varies from $2^1$ to $2^8$, with $PLR=0.15$ and $R=25$, with either a Bernoulli or Gilbert Elliott channel.
The results are plotted in Fig.~\ref{fig:varf}.

The main result is that the two smallest finite fields ($\mathbb{F}_{2}$ and $\mathbb{F}_{4}$) lead to poor performances. Even if the binary field ($\mathbb{F}_{2}$) is attractive because all operations are implemented with extremely fast XORs operations, this field must be avoided in our case. The best compromise seems to be the field $\mathbb{F}_{8}$ which obtains excellent decoding performance while supporting very fast operations. 
The decoding performance differences between $\mathbb{F}_{8}$ and larger finite fields is relatively negligible for both channels. 
This observation remains true for other loss patterns.
We therefore suggest to always use $\mathbb{F}_{8}$.
Additionally \cite{bloemer95xor} explains that a multiplication in the field $\mathbb{F}_{2^m}$ (in our case $m=3$) can be implemented on average with $m/2$ XOR operations per data unit (in our case $3/2$) which can be a useful way of mitigating the processing load of operations over $\mathbb{F}_{8}$.

\section{On the robustness of Tetrys versus FEC block codes in dynamic environments}
\label{sec:easyConfig}
\begin{figure*}[t]
  \begin{center}

  \mbox{
  	\leavevmode
  	\subfigure [PLR=$6$\% and R=$0.2$]
  	{ \label{fig:dstrib_0.06n4}
  	  \includegraphics[keepaspectratio=true, width=\imgratio\columnwidth]{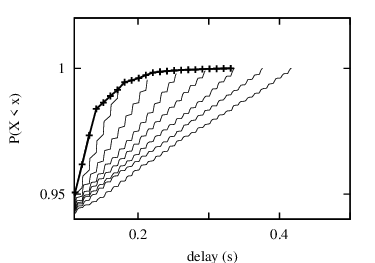} }
  
  	\leavevmode
  	\subfigure [PLR=$9$\% and R=$0.2$]
  	{ \label{fig:dstrib_0.09n4}
	\includegraphics[keepaspectratio=true, width=\imgratio\columnwidth]{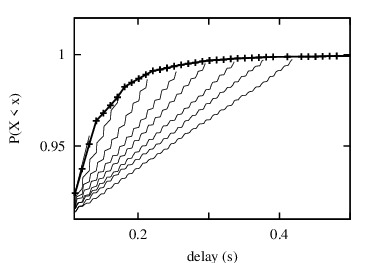} }

  	\leavevmode
  	\subfigure [PLR=$12$\% and R=$0.2$]
  	{ \label{fig:dstrib_0.12n4}
	\includegraphics[keepaspectratio=true, width=\imgratio\columnwidth]{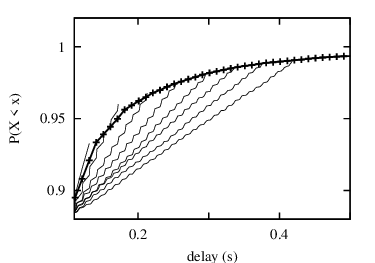} }

  	\leavevmode
  	\subfigure [PLR=$15$\% and R=$0.2$]
  	{ \label{fig:dstrib_0.15n4}
	\includegraphics[keepaspectratio=true, width=\imgratio\columnwidth]{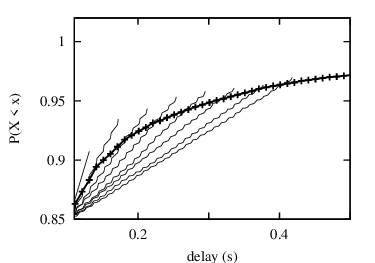} }
}
  \mbox{
  	\subfigure [PLR=$15$\% and R=$0.5$]
  	{ \label{fig:dstrib_0.15n1}
	\includegraphics[keepaspectratio=true, width=\imgratio\columnwidth]{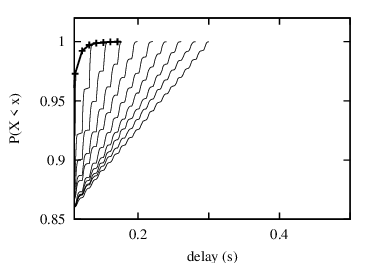} }
  	\leavevmode
  	\subfigure [PLR=$20$\% and R=$0.5$]
  	{ \label{fig:dstrib_0.2n1}
  	  \includegraphics[keepaspectratio=true, width=\imgratio\columnwidth]{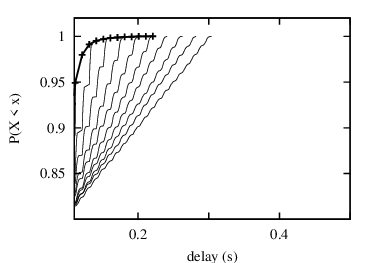} }
  
  	\leavevmode
  	\subfigure [PLR=$25$\% and R=$0.5$]
  	{ \label{fig:dstrib_0.25n1}
	\includegraphics[keepaspectratio=true, width=\imgratio\columnwidth]{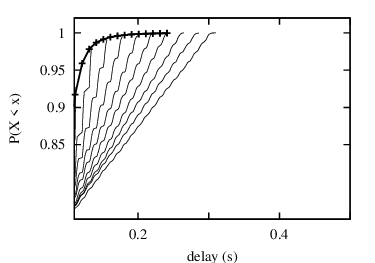} }

  	\leavevmode
  	\subfigure [PLR=$30$\% and R=$0.5$]
  	{ \label{fig:dstrib_0.3n1}
	\includegraphics[keepaspectratio=true, width=\imgratio\columnwidth]{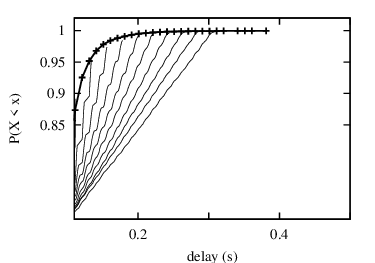} }
  }\\
  
\caption{Cumulative Distribution Functions (CDF) of packets delivery delay for Tetrys (bold curve) and FEC (multiple staircase-like curves, corresponding to various block size configurations), for different packet loss rates and different $R$ values ($0.2$ (upper row) vs. $0.5$ (lower row)).
The RTT is set to $200ms$ and the FEC scheme block size is set to k=\{4; 8; 12; 16; 20; 24; 28; 32\} for the upper row (resp. k=\{2; 4; 6; 8; 10; 12; 14; 16; 18; 20\} for the lower row).}
\label{fig:distribFECTET}
  \end{center}
\end{figure*}


This section compares Tetrys with another usual loss recovery scheme, namely FEC block codes, focusing on the decoding delay metric, a key performance metric with real-time multimedia applications.
In particular, this section emphasizes the simplicity of Tetrys configuration (controlled by a single parameter) and the stability of the performance achieved as the network conditions change.

\subsection{Comparison with FEC block codes}

FEC block codes for the erasure channel are a usual way of mitigating packet losses.
For instance the IETF FECFRAME working group\footnote{See \url{http://www.ietf.org/dyn/wg/charter/fecframe-charter.html}} aims at defining a generic framework between the RTP and UDP protocols to plug various FEC block codes in a very flexible way, to protect one or several application flows, separately or together.
The FEC Framework architecture  being defined \cite{Fecframe.id} is similar to the robust streaming solution that can be found for instance in the 3GPP MBMS or DVB IP Datacasting services \cite{Luby.tob07}.
Rather than focusing on a particular FEC block scheme (e.g. the Raptor codes used in the 3GPP or DVB streaming services \cite{Luby.tob07} or one of the codes considered in \cite{Matsuzono.lcn10}), we consider an MDS FEC code, i.e. a code optimal in terms of correction capabilities.
Note that, even if Raptor codes are often used in streaming services, their rateless feature is totally useless in these environments (\cite{Fecframe.id} (Section 8.1) forbids the code rate to be lower than $0.5$).
Similarly the large block feature of Raptor codes is totally useless in these environments, because of the application real time constraints.

In the remaining of this paper, the term "FEC scheme" will refer to the streaming solution, compliant with the FEC Framework architecture, using an MDS FEC block code. The exact nature of the code is irrelevant, we just know that practical codes will not perform better than the one we are considering in our tests.

This FEC scheme works as follows.
Source packets are sent as soon as the application makes them available.
Then, after the transmission of the $k$ source packets, $n-k$ FEC repair packets are sent (instantaneously).
Since we want to compare Tetrys with the best FEC scheme, we assume that the link bandwidth is sufficiently important to absorb the burst resulting from the introduction of these $n-k$ repair packets.

This approach faces two main limits:
First of all, because of its per-block approach, the recovery of lost packets is only possible at the end, when at least $k$ packets have been received for this block.
This of course introduces a delay that depends on the chosen $k$ parameter: the larger the $k$ value, the better in terms of erasure recovery, but the higher the decoding delay, and the real-time feature of the application anyway incurs an upper limit to $k$.
On the opposite Tetrys repair packets are uniformly spread among source packets.
Therefore lost packets may be recovered without waiting for the end of a fixed length block and without any dependence on the RTT.

Additionally, in real conditions, the PLR is not constant over the time and two key parameters of the FEC scheme, namely the block size ($k$) and the code rate ($k/n$), should be adapted appropriately.
Unfortunately, this adaptation requires feedback information which is, by definition, constrained by the RTT.
Thus, the information is always returned at least one RTT later and might not reflect the current network state.
As a result, the FEC parameters effectively used by the FEC scheme are not necessarily optimal.
On the opposite, Tetrys is controlled by a single parameter and we will show in the following section that it is highly tolerant to varying network conditions.

\subsection{Decoding delay performance evaluation}

We carried out several tests to compare Tetrys to various FEC scheme configurations, i.e. different $k$ and $n$ values, in a Bernoulli channel.
Considering many FEC scheme configurations is important since we do not have any reliable way to identify a priori the best FEC scheme configuration in a given channel.
The results are depicted in Fig.~\ref{fig:distribFECTET}.
The redundancy ratio is set either to $R=0.2$ (i.e. code rate=$0.8$) (upper row) or $R=0.5$ (i.e. code rate=$0.5$) (lower row).
Then, in each figure, there are as many FEC scheme curves as there are possible $k$ values, while keeping the target $R$ (which defines $n$).
The PLR is then progressively increased to approach the $R$ parameter.

For a given code rate we see that in all the studied cases, Tetrys provides full reliability as the CDF tends to one (but this is not the main goal).
This is not the case for the different FEC schemes, essentially with short-dimension FEC codes.
More importantly, the probability for Tetrys to decode below a given delay is higher than most FEC scheme configurations (i.e. the Tetrys curve is higher).
When this is not the case, the FEC scheme features a lower correction capability (i.e. the curve stops earlier and never reaches 1, as in Fig.~\ref{fig:dstrib_0.15n4}).  
However, as the PLR approaches $R$ (e.g. in Fig.~\ref{fig:dstrib_0.15n4}), the Tetrys recovery delay increases and the FEC schemes then overtake Tetrys.

In summary, Tetrys exhibits the same delay and resilience efficiency for most PLR, while being significantly more efficient than the best FEC scheme.
The Tetrys redundancy ratio, $R$, only needs to be dynamically adapted when the PLR increases and be kept sufficiently high compared to the observed PLR.
Since there is a single parameter, this one-dimensional problem is easily addressed.
However we must point out that the main objective in this context is to reduce the recovery delay and not necessarily to optimize the bandwidth occupancy.
An algorithm allowing both a dynamic adaptation of $R$ and the minimization of the bandwidth occupancy will be introduced in Section~\ref{sec:redundancyAlloc}.

\section{Benefits of Tetrys with video-conferencing applications}
\label{sec:video}
\subsection{Specificities of these applications and consequences}
\label{sec:video_intro}

Video-conferencing applications have three main characteristics.
First of all, the end-to-end delay must not exceed 100 ms (see \cite{WengerH264IP03} \cite{TobagiD96}) in order to preserve interactivity.
They are also characterized by their Variable instantaneous Bit Rate (VBR). 
Indeed, Intracoded frames (I-frame), because they are coded from scratch, generate more data than predicted coded frames (P-frames), and even more than bipredicted frames (B-frames).
Finally, losing an I-frame has, in general, a worse impact on the experienced video quality than losing a P or B-frame.

This has several impacts. First of all, FEC schemes are limited by their block size which must neither be too large (since it would impact the end-to-end decoding delay) nor too small (since it would reduce the robustness in front of loss bursts).
Using both the optimal block size and redundancy ratio requires an intricate adaptation mechanism.
On the opposite, Tetrys offers, as seen in Section \ref{sec:easyConfig}, a better compromise between the decoding delay and the resilience than the best FEC scheme.

In the presence of VBR sources as video, this behavior is furthermore confirmed as FEC schemes lack adaptability compared to Tetrys. Indeed, recovering from a given number of losses means waiting for the reception of (at least) the same number of repair packets. With Tetrys, since two consecutive repair packets are spaced with $k$ source packets, when the instantaneous packet rate increases during the transmission, the time needed to receive additional repair packets is reduced, and the probability to recover losses before the deadline increases.
With video coded data, I-frames are the ones that will benefit the most from the adaptability of Tetrys. Although it could be considered only as a side effect of the Tetrys mechanism, this particularity has a major impact on the end user quality as the I-frames have the biggest weight in the video quality measure.

In this sense, Tetrys acts as an Unequal Erasure Protection (UEP) scheme such as DAUEP \cite{DAUEP} or PET \cite{PET}.

More generally, nothing would prevent the use of UEP schemes embedded in Tetrys just by allocating lower code rates to the set of important data or by nesting sources subsets. Hence, in this work we do not consider  any of the FEC UEP schemes nor the Tetrys UEP schemes and let these aspects for a future work.

\subsection{Experimental Setup}

The goal of the tests is to compare Tetrys to various FEC schemes, using either a Bernoulli or GE channel model, during a video transmission.
Various FEC schemes are used, of parameters $(k,n)={(3,4),(6,8),(9,12),(12,16)}$, all of them having the same code rate.
We use the latest ITU-T's video codec recommendation, H.264, and the JM 15.1 H.264/AVC software \cite{h264-software}.
We consider the Foreman sequence, in CIF size, with a frame skip of one picture, resulting in a frame rate of 15 fps.
One I-frame is inserted every 14 P-frames and B-frames are not used at all because of the extra delay B-frames would generate.
The average bitrate is about 384 kbps at the output of the video coder and the coded stream is packed into 500 bytes long packets.
The maximum decoding tolerable delay is set to 100 ms, all the packets received after this due time being dropped.
A total of 150 coded frames, corresponding to 10~seconds of video, is used.
In order to obtain representative results, each sequence is repeated 20 times, leading to the transmission of a sequence composed of 3000 frames and 200~seconds long.
This setup is derived from the common testing conditions mentioned in \cite{WengerH264IP03}.
For evaluating the video we use the Evalvid framework described in \cite{EvalvidKlaueRW03}, where the video quality is measured with the Peak Signal to Noise Ratio (PSNR) metric.

\subsection{Video transmission performance evaluation}

\begin{figure}[ht]
\begin{center}
	\subfigure [Average PSNR with a Bernoulli channel]
  	{ \label{fig:psnr_bs1}
	\includegraphics[keepaspectratio=true, width=\imgratiobis\columnwidth, angle=-90]{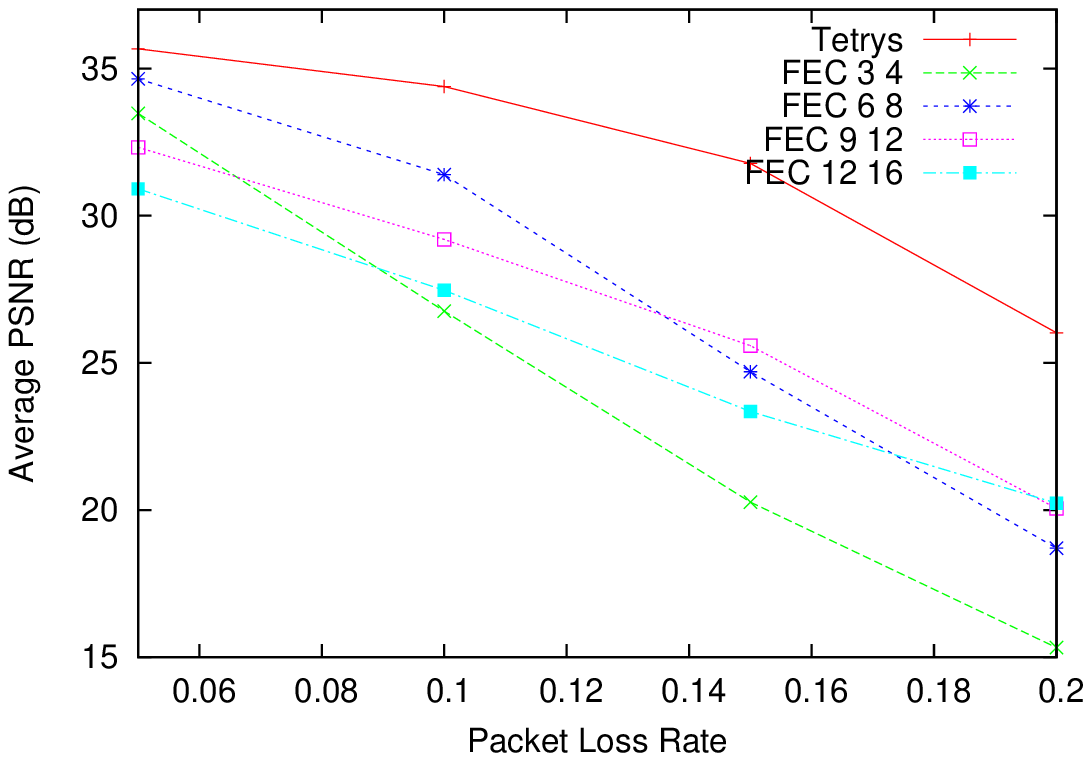}}
  	\subfigure [Average PSNR with a GE channel, burst size=2]
  	{ \label{fig:psnr_bs2}
	\includegraphics[keepaspectratio=true, width=\imgratiobis\columnwidth, angle=-90]{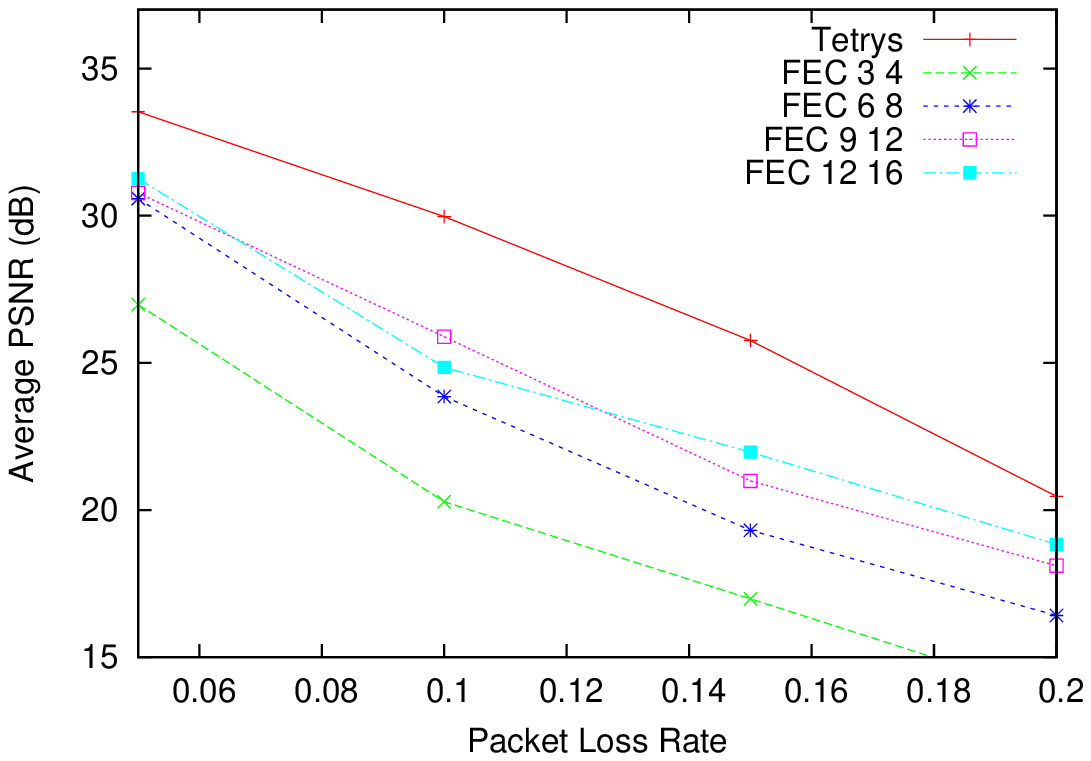}}
  	\subfigure [Average PSNR with a GE channel, burst size=3]
  	{ \label{fig:psnr_bs3}
	\includegraphics[keepaspectratio=true, width=\imgratiobis\columnwidth, angle=-90]{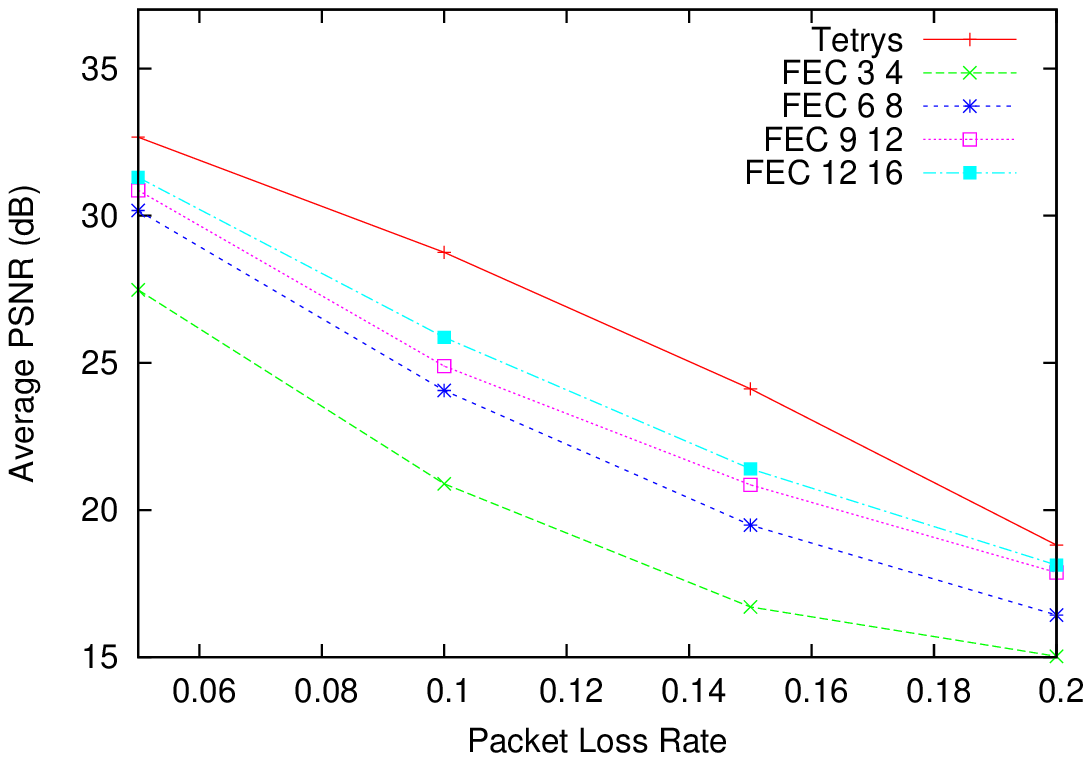}}
\caption{Average PSNR performance of Tetrys versus various FEC schemes during a video sequence transmission, for various channel types.}
\label{fig:psnr_bs}
\end{center}
\end{figure}

\begin{figure}[h]
\begin{center}
\includegraphics[keepaspectratio=true, width=\columnwidth, angle=0]{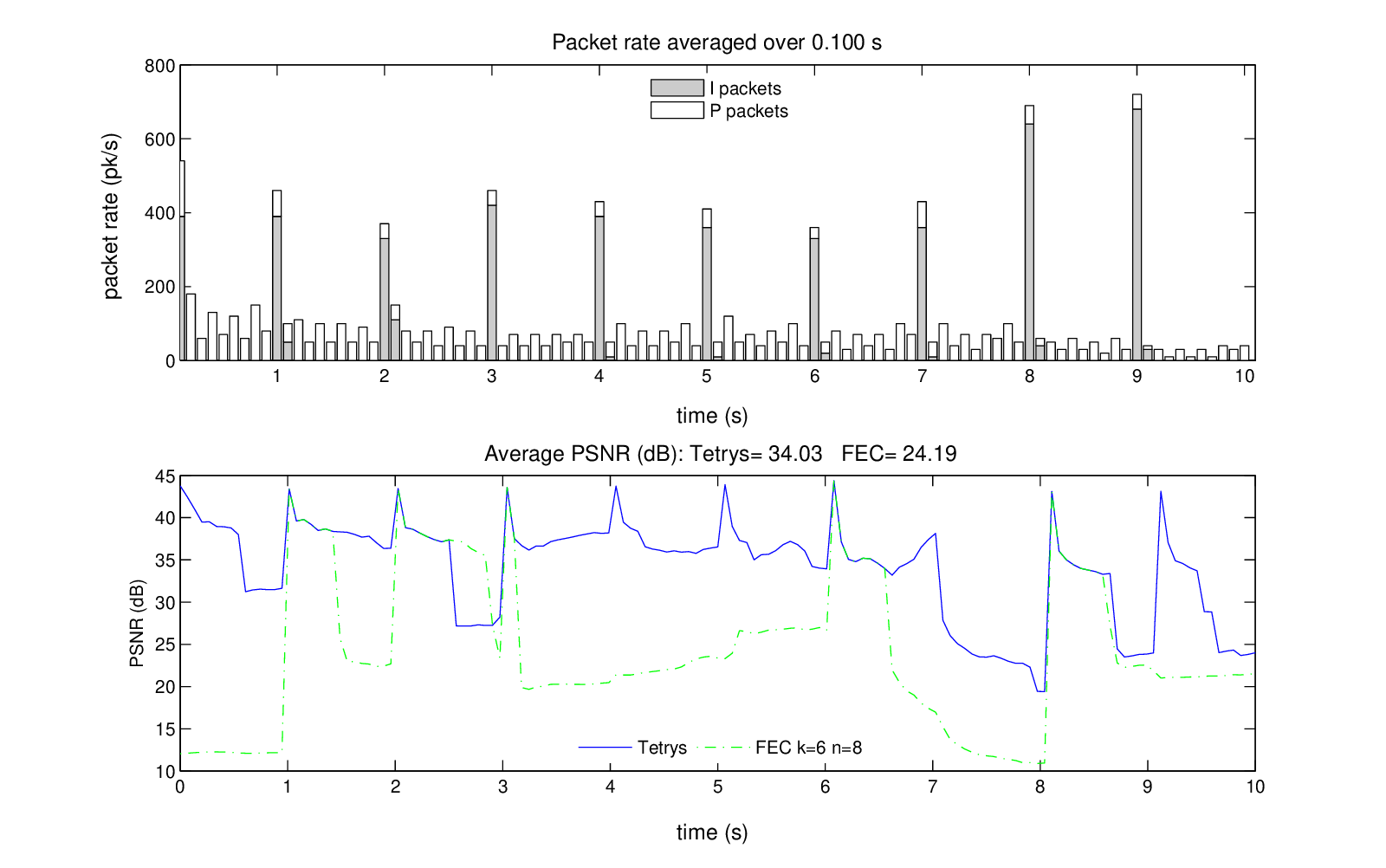}
\caption{Packet rate (top) and instantaneous PSNR of Tetrys versus FEC(6,8) (bottom) during a 10 second snapshot, with a Bernoulli channel and PLR=15\%.}
\label{fig:psnr_instant}
\end{center}
\end{figure}

Let us consider the case of a Bernoulli channel first. Fig.~\ref{fig:psnr_bs1} shows that Tetrys achieves an average PSNR gain of 7.19 dB over the best FEC scheme, namely FEC$(6,8)$ at a PLR of 15\%. The average PSNR drop for Tetrys does not exceed 4 dB when the PLR increases from 5\% up to 16\%, hence ensuring that the average PSNR still remains above 30 dB. When full reliability is impossible because of high time-constraints, Tetrys allows a graceful degradation of the video quality. 
If we consider instantaneous (rather than average) PSNR performances, a representative 10 second trace being shown in Fig.~\ref{fig:psnr_instant}(bottom), Tetrys still outperforms FEC$(6,8)$, the best FEC scheme for this scenario. Tetrys exhibits a significantly higher instantaneous PSNR, except between time $2.5$ and $2.8$, where the FEC scheme behaves momentarily better.
By looking more carefully at the traces over this 10 seconds snapshot (not shown in the figure), we can see that Tetrys retrieved 9 I-frames out of 10, whereas FEC scheme retrieved only 5 I-frames.
This behavior confirms what we said in Section~\ref{sec:video_intro}, namely that I-frames automatically benefit from a better protection compared to P frames with Tetrys. The reason is that Tetrys allows the use of
more redundancy packets in the decoding process before the 100 ms than FEC which is constrained by its block size.
As a matter of fact, if the FEC parameters were adapted with an oracle (instantaneoulsy and automatically), we should obtain similar performance than Tetrys (See Section \ref{sec:easyConfig} for further details.).
This UEP-like behavior is achieved transparently by Tetrys, without requiring any extra information exchange (data types, sizes, or importance) from the source coding application, whereas most of the existing UEP schemes do.

%
%

Let us now consider the case of the GE channel. The average PSNR performances, plotted in Fig.~\ref{fig:psnr_bs2} and \ref{fig:psnr_bs3}, show the same tendency even if the gains are less important: Tetrys still offers a 3.78~dB gain for burst length of 2 and 2.72~dB gain for burst length of 3 over the best FEC scheme.

Therefore, the results achieved are unequivocal: Tetrys clearly outperforms all the tested FEC schemes in all the scenarii, in particular because of its transparent UEP-like behavior with video flows.

\section{Redundancy allocation in Tetrys under reliability and latency constraints}
\label{sec:redundancyAlloc}
As for the video conference example, rather than full reliability, some multimedia applications require that a given proportion $Pkt_{min}$ of packets arrive within a tolerable delay $D_{max}$ (e.g. VoIP applications). After this delay, packets are considered as lost by the application although they might be delayed in the network and arrive later. 

In order to verify whether the request given by an application defined by $(Pkt_{min}, D_{max})$ is feasible, we choose to infer 
a Tetrys heuristic model~$\theta$ following several experiments. We define this model as follows: 
\begin{equation}
\theta(t)_{(d,p,b,T,R)}
\end{equation}
This model gives the cumulative distribution function of the lost packets recovery delay where $R$ is the redundancy ratio for an application that produces a packet every $T$ seconds\footnote{We assume a Constant Bit Rate (CBR) where all the packets have the same size.} 
according to the network characteristics (i.e. a delay $d$, a PLR $p$ and a burstiness of losses $b$).

We then test the capability of Tetrys to satisfy the request $(D_{max}, Pkt_{min})$, given $R$, with a boolean function denoted $\Psi_{\theta(t)}(D_{max},Pkt_{min})$. $\Psi$ returns TRUE if the probability that a packet arrives before $D_{max}$ is higher than $Pkt_{min}$ and FALSE otherwise.
As a result, by iterating $R$ (starting from $R=p$), we find the set of solutions that satisfies the application requirements.
Finally, among these possible solutions, the Tetrys sending application solves Equation~(\ref{eq:tetrys}) to find the smallest redundancy ratio needed denoted $R_{min}$:

\begin{equation}
R_{min} = min (R | \Psi_{\theta(t)}(D_{max},Pkt_{min}))
\label{eq:tetrys}
\end{equation}

The following sections detail the method used to build this model.

\subsection{Model of the delay distribution}

The behaviour of the Tetrys mechanism can be modeled by a Markov chain process with a random walk driven by the losses of source packets and the reception of redundancy packets.
As in Section \ref{subsec:decodingdelay}, we could compute the recurrence and hitting times of the Markov chain and obtain an analytical model of $\theta$.
Unfortunately, the computational complexity of this model requires substantive computation time and prevents any implementation inside a real protocol. 
This motivates the use of our heuristic model $\theta$ previously introduced.

\subsubsection{Experimental setup}

We have performed several experiments with a redundancy ratio $R$ ranging from $0.1$ to $0.5$, a PLR $p$ ranging from $1$\% to $50$\% which follows either a Bernoulli model or a GE model with an average burst size of $2$ or $3$.
For each experiment, $10^5$ source data packets are generated.

\subsubsection{Distributions fitting}

We seek to estimate the delay in number of packets sent (and supposed to be received) between a lost packet and the redundancy packet that rebuild it.
Following the distribution of packets recovery delay obtained by the experiments, we find out that the Weibull law fits our distribution\footnote{We used \texttt{R} \cite{soft:R} statistical software environment}.

A Weibull distribution is defined by two parameters: the scale and the shape. Such distribution captures both exponential distribution if the shape parameter $\kappa$ is around $1$ and the heavy tailed distribution if $\kappa < 1$ and is defined as follows: 
\begin{equation}
P[X<x]=1-e^{-(x/\lambda)^\kappa}
\end{equation}

\subsection{Estimating the distribution parameters}
\begin{figure}[h]
\begin{center}  
\includegraphics[ width=9cm]{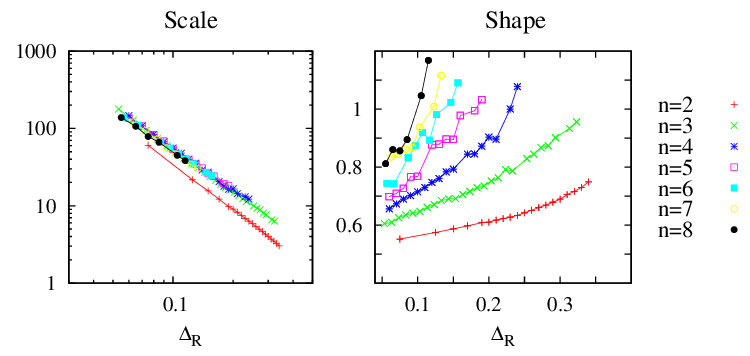}
\end{center}
\caption{Evolution of the scale ($\lambda$) and ($\kappa$) shape as a function of ${\Delta}_{R}$}
\label{wei:ber}
\end{figure}

For a given loss distribution (e.g. Bernoulli or Gilbert-Elliott) the delay distribution is impacted by $n$ ($n=k+1$) and $p$ (as ${\Delta}_{R}=\frac{1}{n}-p)$). For each value of the block size $n$ and each 
loss distribution the shape parameters evolves ``linearly'' as a function of ${\Delta}_{R}$ as seen in Fig.~\ref{wei:ber}. The linear function coefficients obtained through a least square are stored 
in table \ref{tab:shape}.

In the same way, the scale parameter is only impacted by $n$ and the losses distribution. 
The scale can be approximated by:
\begin{equation}
\lambda({\Delta}_{R})=\frac{a}{{{\Delta}_{R}}^b}
\end{equation}
with $a$ and $b$ some parameters related to the loss pattern and $n$. 

It results that $\theta$ can be approximated by:
\begin{equation}
1-e^{-(\frac{x}{\lambda(n,p,c)})^{\kappa(n,p,c)}}
\end{equation}
with:
\begin{itemize}
\item $\lambda(n,p,c)=\frac{a_{c,n}}{(\frac{1}{n}-p)^{b_{c,n}}}$,
\item $\kappa(n,p,c)=a_{c,n}*(\frac{1}{n}-p)+b_{c,n}$,
\item $c$ the channel $\in {ber,b2,b3}$, $a_{c,n}$ and $b_{c,n}$ the appropriate values in the table \ref{tab:shape} and \ref{tab:scale}.
\end{itemize}
\begin{figure}[h]
\begin{center}  
\includegraphics{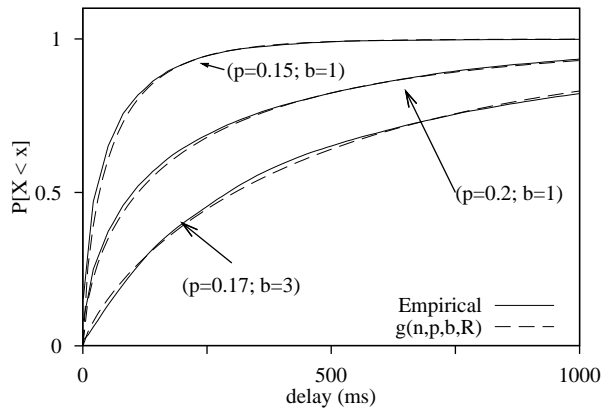} 

\end{center}
\caption{Comparison between the empirical distribution obtain by experiments and $\theta[d,p,b,T,R]$; $T=10ms$, $n=3$.}
\label{wei:example}
\end{figure}

Fig.~\ref{wei:example} presents the good fitting obtained by the empirical distribution of the delay obtained by experimentation and the expected distribution obtained with $\theta$. The results are shown for a PLR of 15\% and 20\% with $b=1$ (i.e. a Bernoulli erasure channel) and a PLR with $b=3$ (a Gilbert-Elliott losses with an average burst size of 3). 

\subsection{Accuracy of the approach}

\begin{figure}[h]
\begin{center}  
\includegraphics[ width=8cm]{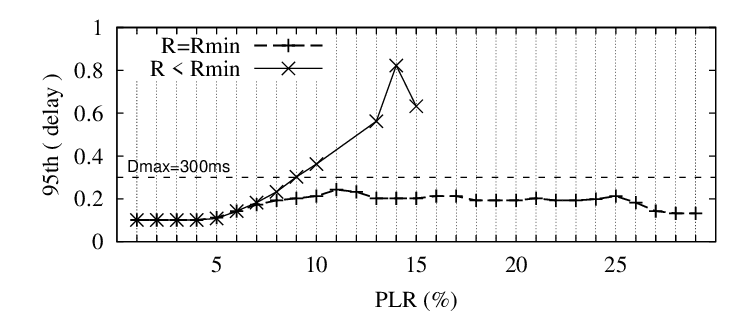}\label{wei:eval:ber}
~\\
\includegraphics[ width=8cm]{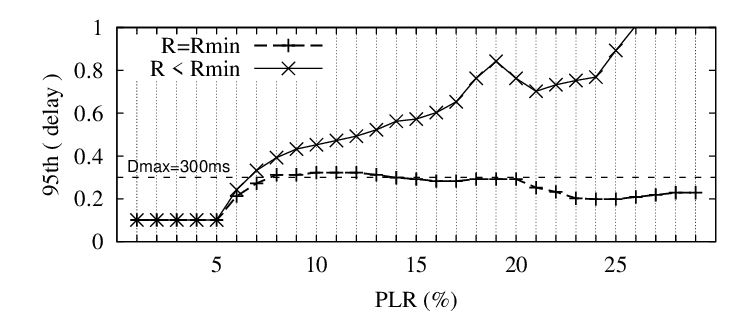}\label{wei:eval:gil}
\end{center}
\caption{Comparison between the optimal (i.e. $=R_{min}$) and suboptimal (i.e. $\leqslant R_{min}$) redundancy ratio for the Bernoulli (top) and GE with average burst size 3 (bottom) models. 
The metric is the $95^{th}$ percentile of the delay.}
\label{wei:eval}
\end{figure}
This mechanism has been implemented and evaluated with the ns-2 network simulator.
Fig.~\ref{wei:eval} shows the results of the accuracy of $R_{min}$ (see (\ref{eq:tetrys})) in a practical use case. The application emits at $100$ pkt/s and requests a minimum of $(Pkt_{min},D_{max})=(0.95,300ms)$ 
and the one-way delay is fixed to $100ms$. 

The figure gives the $95^{th}$ percentile of the delay. According to the application requirements, it should remain below $300ms$. 
Considering a Bernoulli erasure channel, using $R_{min}=\frac{1}{n}$ allows to keep the $95^{th}$ percentile of the delay below $D_{max}$ thus satisfying the application requirements.
When using $R=\frac{1}{n+1}$, the $95^{th}$ delay is higher than $D_{max}$ and does not satify the application requirements.
Considering a Gilbert-Elliott (GE) erasure channel with average burst of $3$, the comparison between $R_{min}=\frac{1}{n}$ and $R=\frac{1}{n+1}$ remains the same. However, we observe when the loss ratio is between 8\% and 12\% that 
the $95^{th}$ percentile of the delay is slightly higher than $D_{max}$. 
The explanation comes from the moving average method used to compute the packet loss rate that sometimes under-estimate this value in the context of GE channel \cite{trafficPLR}.
To conclude, $R_{min}$ is effectively the smallest redundancy ratio compliant with the application requirements.

\section{Conclusion}
\label{sec:conclusion}

In this paper we propose a novel reliability mechanism, Tetrys, based on on-the-fly erasure coding techniques. %
%
We demonstrate, through a detailed modeling of Tetrys performance as well as real measurements, that Tetrys can achieve a full reliability service even in case of an unreliable acknowledgment path (thanks to the non sensitivity of Tetrys to the loss of acknowledgments), or as the extreme case no acknowledgment at all, while ensuring faster data delivery to the application than pure FEC based techniques. In particular, we demonstrate that Tetrys offers key benefits when used in the context of video-conferencing (and more generally real-time applications) over best effort networks. In this case, the main challenge tackled by Tetrys is to combat loss and delay in order to bring a substantial gain in terms of end user perceived quality. We show that Tetrys allows a faster recovery of missing information compared to block codes, and at the same time avoids non-useful retransmitted packets.
Although the contributions of this paper deal with real-time data flows, Tetrys can also be used with non real-time applications,
or at a different protocol layers.
We expect to investigate these considerations, as well as the interactions between Tetrys and a congestion control mechanism, in a future work.


\section*{Acknowledgements}
This work was supported by the French ANR grants 2006 TCOM 019 (CAPRI-FEC project) and ANR-09-VERS-019-02 (ARSSO project).


\appendices

\begin{table}
\begin{center}
\begin{small}
\begin{tabular}{|c|c|c|c|c|c|c|c|c|c|}
\hline
N & $1$ & $2$ & $3$ & $4$ & $5$ & $6$ & $7$ \\ 
\hline
$a_{ber}$ & $0.72$ & $1.25$ & $2.0$ & $2.65$ & $3.44$ & $3.866$ & $5.6$ \\ 
$b_{be}$ & $0.473$ & $0.51$ & $0.512$ & $0.525$ & $0.53$ & $0.55$ & $0.46$ \\
\hline
$a_{b2}$ & $0.48$ & $1.31$ & $1.92$ & $2.15$ & $3.69$ & $5.15$ & $4$ \\ 
$b_{b2}$ & $0.57$ & $0.6$ & $0.61$ & $0.62$ & $0.56$ & $0.48$ & $0.67$ \\
\hline
$a_{b3}$ & $0.62$ & $1.8$ & $2.8$ & $4$ & $4.54$ & $5.5$ & $5.4$ \\ 
$b_{b3}$ & $0.65$ & $0.61$ & $0.57$ & $0.53$ & $0.6$ & $0.62$ & $0.72$ \\
\hline
\end{tabular} 
\end{small}
\end{center}
\caption{Table of linear function coefficients to generate the shape parameter $\kappa$}
\label{tab:shape}
\end{table}

\begin{table}
\begin{center}
\begin{small}
\begin{tabular}{|c|c|c|c|c|c|c|c|c|}
\hline
N & $1$ & $2$ & $3$ & $4$ & $5$ & $6$ & $7$ \\ 
\hline
$a_{ber}$ & $0.83$ & \multicolumn{6}{c|}{$0.35$} \\ 
$b_{ber}$ & $1.815$ & \multicolumn{6}{c|}{$2$} \\
\hline
$a_{b2}$ & $4.2$ & $7.15$ & $9.9$ & $10.48$ & $5.6$ & $2.7$ & $6.3$ \\ 
$b_{b2}$ & $1.14$ & $1.35$ & $1.3$ & $1.3$ & $1.65$ & $1.94$ & $1.57$ \\
\hline
$a_{b3}$ & $11.8$ & $11.4$ & $18.2$ & $9.3$ & $7.1$ & $19.1$ & $36$\\ 
$b_{b3}$ & $1.04$ & $1.44$ & $1.3$ & $1.6$ & $1.7$ & $1.28$ & $1.05$\\
\hline
\end{tabular} 
\end{small}
\end{center}
\caption{Table of linear function coefficients to generate the scale parameter $\lambda$}
\label{tab:scale}
\end{table}

\bibliographystyle{IEEEtran}
\bibliography{biblio}

\end{document}